\documentclass[manuscript]{acmart}
\AtBeginDocument{%
  \providecommand\BibTeX{{%
    \normalfont B\kern-0.5em{\scshape i\kern-0.25em b}\kern-0.8em\TeX}}}

\author{Franklin Mingzhe Li}
\affiliation{%
  \institution{Carnegie Mellon University}
  \city{Pittsburgh}
  \state{PA}
  \country{United States}
}
\email{mingzhe2@cs.cmu.edu}

\author{Akihiko Oharazawa}
\affiliation{%
  \institution{Carnegie Mellon University}
  \city{Pittsburgh}
  \state{PA}
  \country{United States}
}
\email{aoharaza@andrew.cmu.edu}

\author{Chloe Qingyu Zhu}
\affiliation{%
  \institution{Carnegie Mellon University}
  \city{Pittsburgh}
  \state{PA}
  \country{United States}
}
\email{qingyuzh@andrew.cmu.edu}

\author{Misty Fan}
\affiliation{%
  \institution{Carnegie Mellon University}
  \city{Pittsburgh}
  \state{PA}
  \country{United States}
}

\author{Daisuke Sato}
\affiliation{%
  \institution{Carnegie Mellon University}
  \city{Pittsburgh}
  \state{PA}
  \country{United States}
}
\email{daisukes@cs.cmu.edu}

\author{Chieko Asakawa}
\affiliation{%
  \institution{IBM Research, IBM}
  \city{Yorktown Heights}
  \state{NY}
  \country{United States}
}

\author{Patrick Carrington}
\affiliation{%
  \institution{Carnegie Mellon University}
  \city{Pittsburgh}
  \state{PA}
  \country{United States}
}
\email{pcarrington@cmu.edu}

\usepackage{graphicx}
\usepackage{array}
\usepackage{booktabs}
\usepackage{enumitem} 
\renewcommand{\arraystretch}{1.3}

\begin{document}



\title{More than One Step at a Time: Designing Procedural Feedback for Non-visual Makeup Routines}


\renewcommand{\shortauthors}{Li et al.}



\begin{abstract}
Makeup plays a vital role in self-expression, identity, and confidence — yet remains an underexplored domain for assistive technology, especially for people with vision impairments. While existing tools support isolated tasks such as color identification or product labeling, they rarely address the procedural complexity of makeup routines: coordinating step sequences, managing product placement, and assessing the final look with accessible feedback. To understand the real-world process, we conducted a contextual inquiry with 15 visually impaired makeup users, capturing real-time makeup application behaviors and their step-by-step information needs and assessment approaches. Our findings reveal embodied, tactile-first strategies; persistent challenges in blending, symmetry, and assessment; and a desire for honest, real-time, goal-aligned feedback. We also interviewed five professional makeup artists, who reviewed participant makeup videos and provided expert responses to participant-raised questions and assessment practices. We contribute a taxonomy of feedback needs in non-visual makeup, and outline design implications for future assistive systems — emphasizing hands-free, conversational interaction and context-aware, procedural support for expressive and independent beauty practices.
\end{abstract}





\maketitle

\section{Introduction}

Makeup serves as a vital medium of self-expression, self-care, and social participation for millions, with approximately 44\% of Americans incorporating cosmetic products into their daily routines \cite{korichi2008women}. However, among the global population of 2.2 billion people with vision impairments \cite{Blindnes59:online}, makeup remains an underexamined domain—despite its profound personal and cultural importance \cite{li2022feels}. For many blind and low vision individuals, makeup is not just about appearance, but about reclaiming agency, autonomy, and identity. As Lucy Edwards, CoverGirl’s first blind ambassador, powerfully noted: \textit{"When I lost my eyesight, I mourned the inability to see myself. Applying makeup became my way of reclaiming control over how I present to the world"} \cite{Blindnes56:online}.

Visually impaired individuals performing makeup routines must navigate a complex series of interconnected steps—from preparation and product selection to applying makeup and self-assessment— that are highly visual \cite{li2022feels}. This challenge spans multiple stages and modalities, impeding full engagement with makeup as a form of self-expression \cite{pradhan2021inclusive,li2022feels}. While people with vision impairments often develop adaptive strategies and highly personalized techniques \cite{BlindGir58:online}, the inherently visual demands of cosmetics—such as color matching, symmetry, and blending—continue to create systemic barriers that are not well-understood.

Efforts by industry and assistive technology providers have begun to address certain isolated aspects of the makeup experience. These include tactile packaging features (e.g., braille labels by L’Occitane and Procter \& Gamble) \cite{Beautyis6:online,Inclusiv52:online}, real-time visual assistance through platforms like BeMyEyes \cite{BeMyEyes41:online} and Aira \cite{HomeAira84:online}, and apps for color detection and matching \cite{mascetti2016towards} and verifying specific makeup steps \cite{EstéeLau26:online}. However, these solutions focus primarily on discrete interactions rather than the full procedural workflow involved in makeup application. As a result, critical gaps remain in understanding how people with vision impairments organize their makeup routines, handle product quantities, identify and correct mistakes, and perform outcome evaluations—all without direct visual confirmation \cite{li2022feels}. Understanding these multi-step, procedural behaviors is critical for informing the next generation of assistive technologies.

To advance equitable access to cosmetic practices, we must first develop a grounded understanding of these procedural behaviors and experiential challenges. Our study responds to this need by examining three core research questions:

\begin{itemize} 
    \item RQ1: How do visually impaired people adopt specific procedures and strategies for makeup application? 
    \item RQ2: What metrics and mechanisms do visually impaired people currently use to evaluate makeup success?
    \item RQ3: How can feedback and interaction methods be improved to enhance makeup application experiences?
\end{itemize}

To investigate these dimensions, we conducted a two-phase study comprising a contextual inquiry with 15 visually impaired individuals experienced in makeup application, followed by semi-structured interviews with 5 licensed professional makeup artists. During the contextual inquiry, we observed participants’ real-world routines, workspace organization, and the types of questions and support they prioritized throughout the makeup process. Building on participants’ expressed interest in expert feedback, we invited five professional makeup artists to review participant videos and respond to participant-raised questions. This professional review served not only to validate the embodied techniques used by participants but also to surface concrete recommendations for accessible tools, product choices, and procedural sequencing. By bridging real-world routines with expert reflection, our study contributes a rare dual perspective — one that respects user expertise while identifying under-leveraged knowledge from beauty professionals. Our research lays the groundwork for the design of novel, context-aware feedback systems that align with users’ actual workflows and aspirations. This procedural perspective not only bridges a key gap in assistive technology research but also reflects the field’s shift toward holistic, user-centered support for complex, embodied activities. 

From the study, we uncovered procedural behaviors of non-visual makeup in reality, such as the relationship between makeup goals and non-visual makeup process design (Section \ref{Relationship between Makeup Goals and Process Design}), adoption and learning of personalized routines (Section \ref{Adoption and Learning of Personalized Routines}), and detailed procedural steps, behaviors, and product choices embedded within the makeup workflow (Section \ref{Procedural Steps, Behaviors, and Product Choices Embedded within the Makeup Workflow}). Beyond application practices, we also present the assessment practices, feedback gaps, and design opportunities in non-visual makeup, which include assessment approaches by people with vision impairments (Section \ref{Existing Assessment Approaches for People with Vision Impairments}), metrics of assessment among makeup procedures (Section \ref{Metrics of Assessment among Makeup Procedures}), taxonomy of questions and feedback needs for non-visual makeup (Section \ref{Taxonomy of Questions and Feedback Needs for Non-visual Makeup}), and expert reflection on embodied strategies and accessible alternatives (Section \ref{Expert Reflections: Affirming Embodied Strategies and Exploring Accessible Alternatives}), and preferred communication and interaction for assessments (Section \ref{Preferred Communication and Interaction for Makeup Assessment and Feedback}). This work makes three key contributions:

\begin{itemize}
    \item An in-depth account of real-world, non-visual makeup practices, including procedural behaviors and adaptation strategies.
    \item A taxonomy of procedural feedback grounded by user needs through observed challenges and verbalized questions.
    \item Design implications for future assistive systems that support expressive, independent engagement with makeup among blind and low vision users.
\end{itemize}

\section{Related Work}
We review three areas relevant to our study: enabling technologies for people with vision impairments, non-visual approaches to makeup application, and existing makeup technologies.

\subsection{Enabling Technology for People with Vision Impairments}
Assistive technologies have profoundly reshaped how individuals with vision impairments engage with the world, enabling greater independence and participation in everyday life \cite{li2021non,li2022feels,li2023understanding,li2024contextual,li2024recipe,chang2024worldscribe,bigham2010vizwiz,guo2016vizlens,ahmetovic2016navcog,sato2017navcog3}. A wide range of tools—such as screen readers, braille displays, tactile feedback systems, object and color recognition applications, and smartphone-based solutions—support key activities like reading, identifying household items, shopping, and navigating unfamiliar environments \cite{bigham2010vizwiz,guo2016vizlens,neiva2017coloradd,li2017braillesketch,ahmetovic2020recog,brady2013visual,kamikubo2024we,yatani2012spacesense,liu2021tactile,li2025oscar}. Recent innovations, including wearable cameras, AI-powered real-time scene descriptions, and indoor navigation systems like NavCog \cite{ahmetovic2016navcog,sato2017navcog3}, have further supported independent living.

Building on these foundational tools, research in assistive technology has explored increasingly specialized domains, aiming to support more context-specific tasks. For instance, some studies focus on improving access to digital content and education, addressing challenges in web accessibility and media navigation \cite{wang2021revamp,krolak2017accessibility,seo2018understanding,ran2025users,bandukda2024context}. Others apply computer vision to support everyday interactions, such as recognizing objects, interpreting user interfaces, and identifying colors in the environment \cite{guo2016vizlens,bigham2010vizwiz,neiva2017coloradd,kianpisheh2019face}. Physical and tactile design strategies—including 3D-printed augmentations and tactile overlays for touchscreen devices—have also been developed to provide non-visual ways of interacting with tools and information \cite{guo2017facade,he2017tactile}.

Despite this progress, many complex, multi-step activities remain difficult to perform. One such example is applying makeup, which involves a highly procedural, embodied process requiring precise, sequential steps. While existing tools can assist with discrete tasks—such as identifying products or selecting colors—they do not address the complexity of multi-step, procedural activities like makeup routines, which demand coordinated sequencing and continuous assessment \cite{li2022feels,araviiskaia2022recommendations}. This gap presents a critical opportunity: to design assistive systems that move beyond basic information access and instead provide dynamic, step-by-step support, enabling people with vision impairments to complete makeup routines independently and confidently.

\subsection{Non-Visual Approaches to Makeup Application}
Makeup is a deeply personal and culturally significant practice, often intertwined with self-expression, identity, and social norms. It plays a meaningful role in how individuals present themselves in relation to gender, professionalism, and cultural rituals \cite{kennedy2016exploring,jairath1role,dellinger1997makeup}. For many, makeup is not merely cosmetic—it is a form of transformation and empowerment, allowing people to align their appearance with their identity and the expectations of various social settings \cite{li2022feels}. 

This significance extends to people with vision impairments, for whom makeup can be just as important in affirming identity and participating in social life. However, the process of applying makeup presents distinct challenges when visual feedback is unavailable \cite{li2022feels}. Unlike sighted users who rely on mirrors to guide placement, blending, and symmetry, visually impaired individuals often turn to non-visual strategies—such as tactile feedback, spatial memory, consistent routines, and verbal or physical assistance from others—to apply makeup with care and precision \cite{pradhan2021inclusive,li2022feels}. For example, Li et al. mentioned that their participants leveraged video calls with their family members to obtain trusted feedback on the makeup. These adaptive practices are essential to support independence and personal agency in makeup routines. Therefore, these findings motivated us to explore the detailed procedures of makeup applications to uncover the design space for assistive makeup technologies, which is essential to the end-user experiences.

While research has acknowledged the social and identity-related dimensions of makeup use by people with vision impairments, it has largely overlooked the procedural realities of non-visual application. Prior studies have highlighted how makeup contributes to expressions of gender, confidence, and professionalism \cite{pradhan2021inclusive,kennedy2016exploring,jairath1role,dellinger1997makeup}, and how individuals navigate social norms, public spaces, and assistance from others when engaging in beauty practices \cite{pradhan2021inclusive,shinohara2011shadow,li2021choose}. Yet, far less attention has been given to the embodied, step-by-step process of applying makeup without vision—how users plan and execute routines, manage application without mirrors, and assess results through tactile or contextual feedback \cite{li2022feels,araviiskaia2022recommendations}. This lack of procedural insight reflects a broader limitation in HCI and assistive technology research, which has often focused on object recognition or static accessibility solutions. Such tools rarely support complex, dynamic tasks like makeup application, which require sequential action, spatial awareness, and continuous feedback. As a result, current assistive technologies fall short in addressing the full scope of challenges faced by visually impaired individuals in this context.

To address this gap, our work explores how people with vision impairments approach the procedural aspects of makeup. We aim to surface their strategies, adaptations, and support needs—insights that are critical for designing accessible technologies that respect both the functionality and cultural meaning of makeup. By foregrounding the embodied, hands-on nature of makeup application, we seek to inform the next generation of assistive tools that move beyond passive description to offer step-by-step, context-aware guidance tailored to real-world use.

\subsection{Existing Technology for Makeup}
Technologies designed to support makeup application have primarily been developed with sighted users in mind. These systems often leverage computer vision and facial recognition to offer personalized makeup recommendations based on skin tone and facial features \cite{jain2009imaging,nguyen2017smart,ou2016beauty}. Others support creative exploration through interactive design tools that use 3D face modeling, tangible interfaces, and projection mapping \cite{treepong2018makeup,kao2016chromoskin}. While innovative, these systems assume a certain level of visual engagement, making them largely inaccessible to users with vision impairments.

Instructional resources such as online makeup tutorials are also widely used by sighted individuals to learn techniques and styles. However, these resources often lack accessibility features such as verbal step-by-step descriptions, content-based voice navigation, or tactile demonstrations \cite{truong2021automatic,chang2021rubyslippers}. Even efforts aimed at supporting marginalized groups—for example, makeup systems tailored for transgender individuals \cite{chong2021exploring}—typically rely on visual interactions, limiting their applicability for blind and low-vision users.

While some aspects of the makeup process—such as product identification or color detection—can be partially addressed using existing assistive technologies, these solutions do not provide the comprehensive, procedural support needed to enable independent makeup application. To fill this gap, we conducted a study with both visually impaired makeup users and professional makeup artists. This collaborative process uncovered a taxonomy of non-visual strategies and feedback needs, laying the groundwork for the development of assistive systems that support the full spectrum of makeup activities—from planning and preparation to application and assessment.

\begin{table*}[]
\Large
\resizebox{\textwidth}{!}{%
\begin{tabular}{p{0.7cm}|p{0.5cm}p{1.2cm}p{9cm}p{2.8cm}p{4cm}}
\hline
\textbf{PID} & \textbf{Age} & \textbf{Gender} & \textbf{Vision Condition}                                                                                 & \textbf{Blindness Onset} & \textbf{Makeup Frequency}      \\ \hline
P1  & 37  & Female & Legally blind left eye, totally blind right eye. & 12 years old    & 2 to 3 times a week   \\
P2  & 25  & Female & Legally blind, brain injury. & 15 years old    & 3 times a week        \\
P3  & 50  & Female & Light perception only. & Since birth     & 3 to 4 times a month  \\
P4  & 65  & Female & Totally blind. & 20 years old    & 1 to 2 times a week   \\
P5  & 23  & Female & Minimal light perception, legally blind. & Since birth     & Once a few months     \\
P6  & 62  & Female & Legally blind, left 20/1000, right 20/800, no central vision. & 55 years old    & Daily                 \\
P7  & 27  & Female & Light perception left eye, totally blind right eye. & Since birth     & Once every two months \\
P8  & 42  & Female & Legally blind, nerve damage. & 25 years old    & Daily                 \\
P9  & 27  & Female & Totally blind, infection. & Since birth     & 2 to 3 times a week   \\
P10 & 43  & Female & Legally blind with light perception, no functional vision. & 17 years old    & 5 to 8 times a month  \\
P11 & 30  & Female & Totally blind, born premature. & Since birth     & Once a couple week    \\
P12 & 33  & Female & Totally blind, congenital. & Since birth     & A few times a week    \\
P13 & 42  & Female & Totally blind, glaucoma and ocular vasculitis. & Since birth     & 3 to 4 times a week   \\
P14 & 36  & Female & Legally blind, congenital glaucoma. & Since birth     & Once a week           \\
P15 & 50  & Female & Legally blind, no central vision, asthma. & 47 years old    & Daily                 \\ \hline
\end{tabular}%
}
\caption{Demographic information of participants with vision impairments.}
\label{table:participants}
\end{table*}

\section{Makeup Contextual Inquiry and Expert Feedback Study}
Our study employed a contextual inquiry approach to explore the real-world practices, preferences, and challenges of makeup application among people with vision impairments. To complement these insights and gather expert perspectives, we also conducted follow-up interviews with licensed professional makeup artists who reviewed participant videos and provided feedback. Our methodology involved detailed observation of participants' natural routines, think-aloud reflections, follow-up discussion, and expert commentary to inform the design of future assistive systems for non-visual makeup. The entire recruitment and interview process was approved by the Institutional Review Board (IRB).

\subsection{Contextual Inquiry with People with Vision Impairments}

\subsubsection{Participants}
To understand the non-visual makeup procedure in reality and discuss preferences of feedback and support for each makeup step, we conducted a contextual inquiry with 15 visually impaired volunteers who have experience with makeup and cosmetics (P1 - P15) (Table \ref{table:participants}). Although our recruitment form asked for participants of any gender identity, all 15 participants who responded identified themselves as female, with an average age of 39.5 (SD = 12.9), ranging from 23 to 65 years old (Table \ref{table:participants}). 5 of them are totally blind, and 10 are legally blind (Table \ref{table:participants}). Regarding their first time doing makeup, 6 of the participants started before the loss of vision, and the rest were after. Their makeup frequency also range from daily to once every few months. Participants were recruited via social platforms (e.g., Reddit, Twitter, Facebook) and email lists through the National Federation of the Blind (NFB). To participate in our interview, participants had to satisfy the following requirements: 1) be 18 years old or above; 2) be legally blind (i.e., visual acuity less than 20/200); 3) have experience with makeup. The interviews were conducted virtually via Zoom, and each one took around 90 minutes. Participants were compensated with a \$60 Amazon gift card for completing the study.


\subsubsection{Study Description}
This contextual inquiry was designed to elicit rich, in-situ insights into the procedural behaviors, support needs, and adaptation strategies of people with vision impairments when applying makeup. Each session was conducted remotely via Zoom and structured to balance observational data collection with reflective discussion. The study consisted of five stages: technical setup, background gathering, pre-application reflection, live makeup application with concurrent verbalization, and post-application debrief. This structure allowed us to capture not only participants’ real-time decision-making and tactile strategies but also their broader experiences, expectations, and unmet needs. Below, we detail each segment of the session.

\textbf{Setup and Technical Check (5 minutes)}: The session began with a setup and technical check to ensure optimal conditions for recording their makeup applications. We verified that Zoom settings, lighting conditions, and camera angles provided clear visibility of the participant's face and makeup application process.

\textbf{Demographic Information and Makeup Background (10 minutes)}: Next, we collected participants' demographic information. This included inquiries about their age, gender, vision condition, and the onset of blindness. We also asked about the frequency and context in which they applied makeup, the types of products they typically used, and their general goals for wearing makeup.

\textbf{Pre-Makeup Reflection (5 minutes)}: Participants were asked to reflect on their approach to makeup application. They described their goal for the current makeup session, whether it was a natural, bold, or professional look. We discussed any specific routines or step-by-step methods they typically followed and how they developed these routines over time.

\textbf{Makeup Application and Think-Aloud Process (30 minutes)}: During this phase, our participants applied makeup as they normally would while verbalizing their thoughts. They described each action they took, explaining their reasoning behind specific steps and any challenges encountered along the way. They also expressed what information or feedback they wished they had during the process, particularly regarding product-related challenges such as applying mascara evenly or ensuring foundation coverage.

\textbf{Follow-Up Discussion (40 minutes)}: After completing the makeup application, participants were asked to reflect on their experience step-by-step. They described their reasoning behind the chosen steps and techniques, the challenges they faced, and how they typically addressed these difficulties. Additionally, we inquired about any feedback they would seek from a professional makeup artist for specific steps. Participants also shared their thoughts on how assistive technology could support their makeup process, as well as the strategies they used to assess their final look and any difficulties in this assessment process.

\begin{table}[]
\large
\resizebox{1\columnwidth}{!}{%
\begin{tabular}{p{0.7cm}|p{0.7cm}p{1cm}p{3.5cm}p{8cm}p{2.5cm}}
\hline
\textbf{PID} & \textbf{Age} & \textbf{Gender} & \textbf{Year of Professional Makeup (licensed)} & \textbf{Makeup Specialization} & \textbf{Makeup Videos Reviewed} \\ \hline
M1 & 25 & Female & 4  & Everyday and natural makeup, bold and dramatic styles, and educational or consultative support            & P5, P9, P15  \\
M2 & 34 & Female & 8  & Practical everyday makeup with emphasis on evaluation, routine guidance, and personalized recommendations & P3, P11, P13 \\
M3 & 25 & Female & 7  & Adaptive and occasion-based makeup for events, with experience in bridal, competitive, and general styles & P6, P12, P14 \\
M4 & 25 & Female & 10 & Bridal and family-centered makeup, including maternity shoots and everyday wearable looks                 & P2, P7, P10  \\
M5 & 25 & Female & 1  & Natural and everyday makeup, and bridal makeup                                                            & P1, P4, P8   \\ \hline
\end{tabular}%
}
\caption{Demographic information of licensed professional makeup artists}
\label{tab:makeupartists}
\end{table}

\subsection{Semi-Structured Interview and Makeup Reflection with Licensed Professional Makeup Artists}
After recording the non-visual makeup applications from people with vision impairments and their questions regarding each individual step (before, during, and after). Given people with vision impairments naturally consult professional makeup artists for color choices, makeup application, and tool recommendations \cite{li2022feels}, we formatted the data collected and presented it to licensed professional makeup artists to facilitate a collaborative exchange of insights. Our aim was to understand how professional perspectives might complement and support the agency of visually impaired users, rather than override their established strategies.

\subsubsection{Participants}
We further conducted a semi-structured interview with 5 licensed professional makeup artists (M1 - M5). Although our recruitment form asked for participants of any gender identity, all 5 participants who responded identified themselves as female, with an average age of 26.8 (SD = 4.0), ranging from 25 to 34 years old (Table \ref{tab:makeupartists}). They have an average of 6 years for professional makeup experiences (licensed). Participants were recruited via social platforms (e.g., Instagram) (Table \ref{tab:makeupartists}). To participate in our interview, participants had to satisfy the following requirements: 1) be 18 years old or above; 2) have professional makeup experience. The interviews were conducted virtually via Zoom, and each one took around 70 minutes. Participants were compensated with a \$40 Amazon gift card for completing the study.



\subsubsection{Study Description}
This study aimed to gather insights from professional makeup artists on assessing and providing feedback on the makeup application process of people with vision impairments. The study was structured into two key phases: demographic inquiry and a combined video review and interview session.

\textbf{Demographic Information (5 minutes)}: To understand the professional background of the makeup artists, we collected demographic information, including their age, gender, years of experience in makeup application, and approaches to handling different makeup needs for diverse clients.

\textbf{Video Review and Interview Session (65 minutes)}: Participants reviewed recordings of people with vision impairments applying makeup. While watching the videos, makeup artists engaged in a structured discussion, responding directly to participants' questions and reflecting on each participant’s personalized approaches. The interview encouraged makeup artists to share suggestions as supportive options, respecting the expertise and routines developed by users themselves. They assessed the quality of makeup application in each video clip, discussed standard practices for verifying and checking makeup, and offered specific recommendations for improvement, with the intention of expanding, rather than constraining, the participants’ strategies.

\subsection{Data Analysis}
The contextual inquiries were meticulously documented using both video recording methods and transcripts of the think-aloud process. For the analysis of the contextual inquiry videos and responses in the study with participants with vision impairments, we used thematic analysis for the transcripts \cite{braun2006using}. Our analysis centered on the three research questions (RQ1 - RQ3). As for the analysis of the semi-structured interviews with professional makeup artists, a similar thematic analysis approach was employed \cite{braun2006using}. This analysis revolved around themes related to feedback to non-visual makeup processes, and recommendations on makeup improvements and enhancement.

\section{Findings: Procedural Behaviors of Non-visual Makeup in Reality (RQ1)}
This section presents key findings on how blind individuals engage with makeup as a procedural, embodied activity. We organize our findings into three subsections: (1) the relationship between makeup goals and process design, (2) the adoption and learning of personalized routines, and (3) the procedural steps, behaviors, and product choices embedded within the makeup workflow.

\subsection{Relationship between Makeup Goals and Process Design}
\label{Relationship between Makeup Goals and Process Design}
In the study, we asked our participants regarding their makeup goal and how does the goal related to makeup processes. Based on the responses, we found our participants leverage different makeup processes align with different goals by context: 1) daily use and everyday life, 2) special occasions and events, 3) professional settings.

\subsubsection{Makeup for Daily Use and Everyday Life} For all participants, makeup was an important part of their daily routine — not for elaborate transformations, but for fostering confidence, maintaining structure, and supporting emotional well-being. In everyday settings, participants favored \textbf{simple, quick-to-apply products} that offered immediate results with minimal effort, such as tinted lip balm, cream blush, or lipstick. These products were often chosen for their ease of use, tactile-friendly formats, and reliability without requiring visual precision. Many participants described daily makeup as a small but meaningful ritual that helped them feel ready to face the day ($N=12$). P2 noted that makeup provided a personal boost of confidence when stepping outside, while P3 shared that even small gestures — like applying lip gloss — created a sense of routine and emotional grounding. When pressed for time, our participants streamlined their process, often relying on just one or two products ($N=8$). Lip products, in particular, were a popular choice for daily wear — offering a quick pick-me-up with noticeable impact and low risk of error (P11, P13). Others, like P7, \textit{"I like to feel natural, like I’m just enhancing what’s already there"}, preferred products that subtly enhanced natural features rather than conceal them, such as skin-tone-matching powders or highlighter sticks.

\subsubsection{Makeup for Special Occasion and Events}
Special occasions prompted a shift in both the intent and complexity of makeup routines. Participants described \textbf{using a wider range of products, often with bolder colors or more elaborate application} for events such as weddings or parties, and often \textbf{require additional help and feedback from sighted people} ($N=9$). P6 clearly distinguished between routine and event-based makeup: \textit{“If I’m just going to the supermarket, I do it alone. But if it’s a wedding, I’ll ask for help—I want it to look perfect.”} For these occasions, she used \textit{foundation, concealer, and bold lipstick}, and often relied on another person to assist with application. P4 discussed changing her makeup style depending on the setting: \textit{“For parties, I use more bold colors, especially around my eyes. I want it to say something.”} This included vibrant or shimmering eyeshadows, which she avoided in everyday contexts. These choices were influenced not only by the formality of the event but also by a desire for self-expression and celebration.

\subsubsection{Makeup for Professional Settings}
In professional settings, such as working or attending conferences, our participants used \textbf{subtle and durable products to convey a polished, competent, and age-appropriate appearance} ($N=7$). For example, P5 said, \textit{“It’s about presenting the best of myself, especially at work or formal meetings,”} and used foundation and neutral-toned products to achieve a consistent, composed appearance. P1 also emphasized the importance of looking professional and used foundation to conceal acne and scars, helping her feel more confident in work environments. P14 shared, \textit{“I want to look my age, not younger or older—just competent and well-kept,”} and chose skin-tone-matching foundation and soft blush to align with that goal. Several participants noted that professional occasions often demand a higher level of visual precision in makeup application to avoid unwanted attention or appearing unkempt. Unlike social or home environments, opportunities to receive sighted assistance or informal feedback are often limited, making self-reliant and precise routines especially important for these settings.

\subsection{Adoption and Learning of Personalized Routines}
\label{Adoption and Learning of Personalized Routines}
We further asked our participants regarding the process of makeup routine adoption, we found four key main themes that are unique to non-visual makeup: 1) learning pathways and social support, 2) trial-and-error and gradual mastery, 3) obtaining product information from professional makeup artists, and 4) learning from media and translating visual content.

\subsubsection{Learning Pathways and Social Support}
\begin{figure}[t]
    \centering
    \includegraphics[width=0.7\columnwidth]{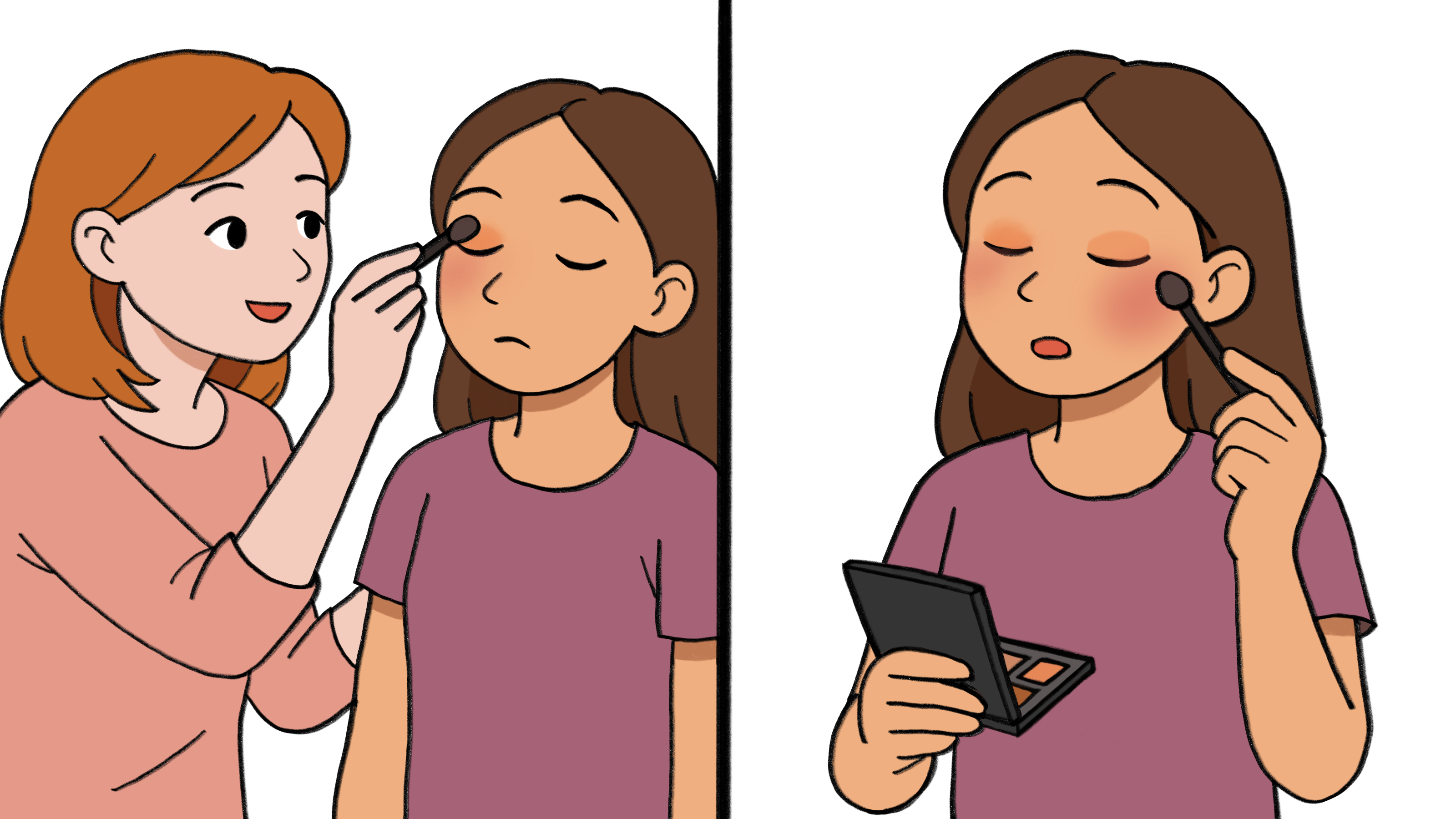}
    \caption{Demonstration of mirrored, side-by-side learning approach. The left figure shows a makeup artist apply makeup for the blind person on the left face. The right figure shows the blind person is trying to replicate makeup for her right face.}
    \label{fig:learning}
    \Description{A two-panel color illustration showing a makeup learning scenario. The left panel shows a makeup artist applying eyeshadow on the left side of a young woman’s face while the woman keeps her eyes closed. The right panel shows the same woman independently applying makeup to the right side of her face, holding an eyeshadow palette. The left side of her face, done by the makeup artist, appears more refined, while the right side, done by herself, has bolder and more uneven blush and eyeshadow application.}
\end{figure}

Our participants often began learning makeup through trusted people, including friends, family members and professional makeup artists, using mirrored practice and feedback to build initial confidence ($N=15$). For example, P2, P5, and P9 all began by observing and listening to relatives describe what they were doing. P3 shared a particularly hands-on approach: \textit{“My friend would do one side, then I’d try the other side myself,”} allowing her to build symmetry through tactile mimicry. Similarly, our participants mentioned they also consult professional makeup artists from beauty stores like ULTA or Sephora to obtain new product information, learn new makeup and skills, and get feedback on new looks. P4 learned from a professional makeup artist who applied makeup to one side of her face, and then invited her to replicate it: \textit{“I’d get feedback on what worked.”} This \textbf{mirrored, side-by-side learning technique—with real-time feedback} (Figure \ref{fig:learning})—emerged as an effective way for participants to internalize motions and tools, particularly for building confidence in unfamiliar tasks.

\subsubsection{Trial-and-Error and Gradual Mastery}
Many participants framed their learning as a process of experimentation ($N=11$). P2 and P6 described a step-by-step approach, beginning with easy routines and adding more complexity. P15 provided a striking example of perseverance: \textit{“Trial and error every day—you find what fits you, you learn it.”} Several participants highlighted the need for independence ($N=7$), especially when others weren’t available. P6 said, \textit{“It’s important for me to be able to do it myself—people can’t always help.”} Others adapted their routines around what they could do reliably on their own. For instance, P1 simplified complex looks: \textit{“Eye makeup is hard—I can’t do smoky eyes anymore. I overblend to make up for skipping primer and foundation.”} This stage was marked by \textbf{building personal routines through tactile cues, memory, and adjustment, reinforcing the importance of adaptability and confidence in solo practice}.

\subsubsection{Obtaining Product Information from Professional Makeup Artists}
To build sustainable and accessible makeup routines, our participants actively sought product information and recommendations from professional makeup artists and beauty consultants ($N=9$), often supplementing their own trial-and-error experiences. Choosing accessible, non-irritating, and versatile products was critical for developing routines they could reliably perform. Specifically, our participants approached product consultations analytically — P10 asked detailed questions about product use, skin sensitivity, and application techniques: \textit{“What’s the intention of this product? How do I remove it safely? What’s the motion I need to take?”} Furthermore, our participants remained cautious about recommendations that did not align with their non-visual practices, such as products requiring fine visual detail or difficult tools. As P8 cautioned, \textit{“Someone tried to sell me eyeliner that needed a brush — I can’t do that.”} These experiences illustrate how obtaining product information from professionals was not only a practical necessity but also a site of critical decision-making, balancing expert guidance with personal accessibility needs.

\subsubsection{Learning from Media and Translating Visual Content}
Our participants selectively engaged with visual tutorials, focusing on content that described tools, motions, and process steps ($N=6$). P13 described YouTube tutorials: \textit{“What tools they use, what product, how they perform—all of it matters. That changed the game for me.”} She focused on transferable knowledge like tool types and order of operations, rather than visual aesthetics. Similarly, P12 and P14 read reviews and articles to understand what products might work for them. P14 noted, \textit{“If anyone said what tool they used and how they used it, that really jumped out at me.”} These participants transformed visual knowledge into non-visual strategies—a creative adaptation that demonstrates both initiative and resourcefulness.

\subsection{Procedural Steps, Behaviors, and Product Choices Embedded within the Makeup Workflow}
\label{Procedural Steps, Behaviors, and Product Choices Embedded within the Makeup Workflow}
\begin{figure*}[t]
    \centering
    \includegraphics[width=1\columnwidth]{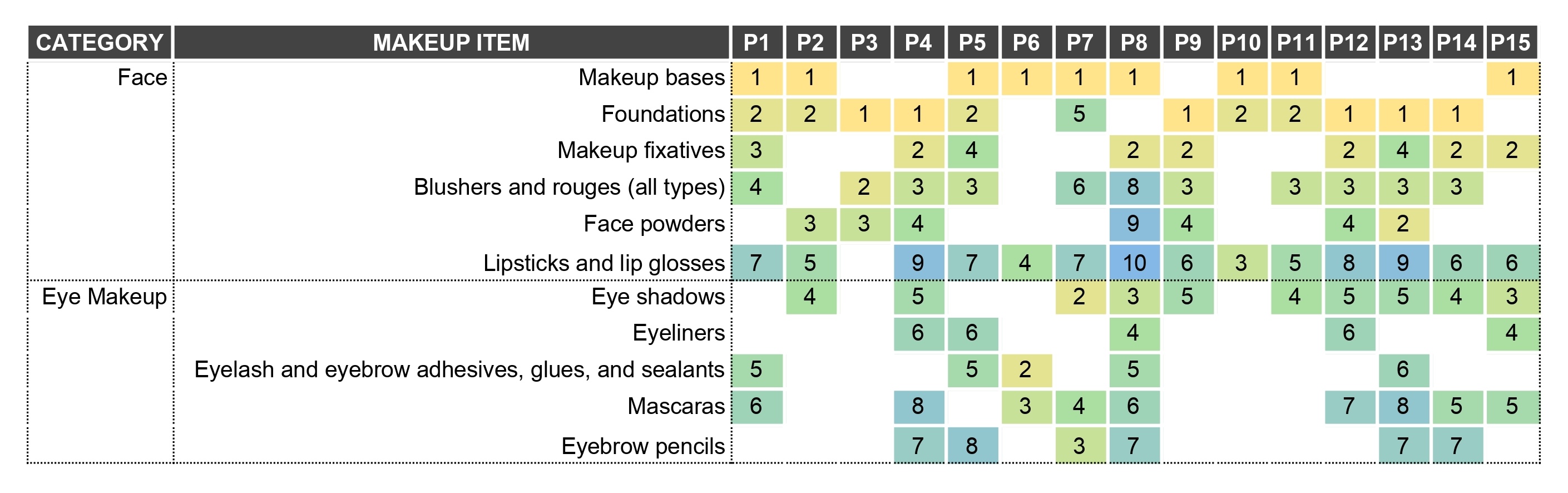}
    \caption{Makeup procedures for non-visual makeup among 15 participants. The number inside each cell indicates the order of the makeup application.}
    \label{fig:procedure}
    \Description{A color-coded matrix showing the makeup steps followed by each of the 15 participants (P1–P15), organized by makeup item and category. Face makeup steps include makeup bases, foundations, fixatives, blush, powders, and lip products; eye makeup steps include eyeshadow, eyeliner, lash and brow adhesives, mascara, and brow pencils. Numbers represent the sequence order of application, with deeper colors indicating later steps. Most participants begin with base and foundation, and end with lips or eye-related products. The chart reveals personalized routines with common patterns across users.}
\end{figure*}

In our study, we had our participants apply their makeup as their daily routine. We then break down their makeup routine through each makeup steps to further understand their procedural approaches (Figure \ref{fig:procedure}). The breakdown of makeup steps based on FDA official document for cosmetic good guidance \cite{us2013draft} as well as the makeup artist handbook \cite{davis2017makeup}. It mainly covered two makeup categories: makeup preparations (i.e., makeup bases, foundations, makeup fixative, blusher and rouges, face powders, lipsticks and lip glosses), and eye makeup preparations (i.e., eye shadows, eyeliners, eyelash and eyebrow adhesives, glues, and sealants, mascaras, and eyebrow pencils) (Figure \ref{fig:procedure}) \cite{davis2017makeup,us2013draft}. In this subsection, we present findings of non-visual makeup behaviors among all makeup procedures (Figure \ref{fig:procedure}).

\subsubsection{Overall Behavioral Patterns and Strategic Decision-Making}
Our participants approached makeup not as a fixed or purely aesthetic routine, but as a flexible, embodied practice shaped by control, sensory awareness, and personal strategy. Their routines reflected a deep understanding of their own capabilities, where every product choice, tool use, and step sequence was carefully considered to balance expression with manageability. Rather than following conventional makeup orders or trends, our participants designed routines that prioritized tactile control, minimized risk, and maximized confidence — demonstrating expertise not only in makeup skills but in navigating the constraints and possibilities of non-visual application.

\subsubsection{Face Products: Foundation, Blush, and Powders as Foundational Steps}
Face products, especially foundation and blush, were essential starting points for most routines — chosen for their tactile qualities and forgiving nature. Our participants favored cream-based foundations for their sensory feedback. P7 appreciated how cream foundation allowed her to feel whether it was properly blended: \textit{“It’s easy to feel if it is blended properly”}. P14 noted the coolness of certain products helped with placement: \textit{“It’s very tactile — I can feel where it is.”} For blush, spatial accuracy was a recurring challenge. P11 commented: \textit{“Blush is harder than eyeshadow — you have to find the exact spot on the cheek.”} To address this, participants like P7 and P13 used their hands to locate cheekbones or tapped the product gently to control dosage and avoid clumping. Loose powders introduced difficulties in dosage estimation, particularly without visual confirmation. P5 and P13 struggled with knowing how much product was applied. To overcome this, P8 used Braille-labeled products and the Seeing AI app for assistance.

\subsubsection{Eye Makeup: Eyeliner, Mascara, and Eyebrows as Selectively Omitted or Adapted Steps}
Eye makeup steps — particularly eyeliner and mascara — were the most frequently omitted parts of participants’ routines, largely due to the precision, dexterity, and risk they demanded.

\textbf{Eyeliner: Precision-Heavy and Often Avoided}.
Only five participants regularly applied eyeliner (Figure \ref{fig:procedure}). For many, the fear of making visible mistakes outweighed the aesthetic payoff. As P5 shared: \textit{“I have no idea if it’s a straight line or not,”} while P7 avoided eyeliner entirely because it caused stress and required skills not worth the effort for everyday looks. Those who did apply eyeliner relied on tactile adaptations. P8 used a technique of pulling the lower eyelid down with one hand while tracing the waterline with the other, building control through practice. P15 favored eyeliner pencils with tactile markings, while P12 applied from the outer corner inward to better match her hand motion.

\textbf{Mascara: Stressful, Risk-Prone, and Often Skipped}. Mascara elicited particular frustration and fear. Participants described the difficulty of aiming the wand, avoiding the eyelid or nose, and gauging product coverage. P6 sometimes had to redo one eye multiple times and resorted to video calls with friends or services like Aira \cite{HomeAira84:online} or FaceTime \cite{FaceTime24:online} to verify results. P12 similarly relied on others for feedback on evenness and smudging. To reduce errors, P7 and P8 preferred miniature mascara wands for better control. P14 recommended using straighter applicators and stabilized handles to avoid poking the eye or over-applying. Others gripped the wand closer to the base for better precision or skipped mascara entirely, particularly for daily wear.

\textbf{Eyebrows: Manageable with Tactile Guidance}.
Eyebrows were generally more approachable but still required care. Those who applied eyebrow pencils, such as P13, emphasized tactile guidance — using fingers to locate the brow bone and brushing hairs upward before outlining with light pressure. This tactile-first method minimized over-drawing and built confidence in placement.

\textbf{Emotional Impact of Omission}.
While step omission was often a strategic choice, it sometimes carried emotional weight. For example, P6 reflected on the loss of certain makeup practices after vision loss: \textit{“When you going on the road, you have to pick your battles. Let it go, move on; for me, it is very upsetting; it reminds me what I cannot do anymore.”}

\subsubsection{Lip Products: A Reliable and Empowering Step Across Routines}
Unlike eye makeup, lipsticks and lip glosses were the most consistently used products across participants (Figure \ref{fig:procedure}) ($N=14$), valued not only for their manageability but also for their emotional significance as the final touch that completed the look and boosted confidence. Our participants described lip application as a highly tactile and predictable process. P15 explained, \textit{“I feel my lip and glide — it’s the easiest thing,”} while others, like P2, used their hands to align the product with the lip edges before applying. Product design also played an important role in supporting precision and control. For example, P14 preferred slender lipstick tubes for their smaller, more error-proof tip: \textit{“It does double duty. I can be more precise, and it’s more error-proof.”} Lip products remained a trusted and empowering staple across everyday routines and special occasions, valued for their tactile nature, ease of correction, and immediate visual payoff.

\section{Findings: Assessment Practices, Feedback Gaps, and Design Opportunities in Non-Visual Makeup (RQ2, RQ3)}
From the observation and follow-up discussion, we identified unique makeup assessment approaches, reasoning for certain steps, and questions that our participants (P1-P15) typically ask. We then talked about responses from professional makeup artists (M1-M5) and preferred designs for feedback and communication.

\subsection{Existing Assessment Approaches for People with Vision Impairments}
\label{Existing Assessment Approaches for People with Vision Impairments}
Our participants employed a wide range of strategies to assess their makeup due to the lack of reliable real-time feedback from existing tools. These strategies included asking others ($N=15$), using general-purpose AI apps ($N=8$), relying on memory or tactile cues ($N=11$), and improvising based on trial and error ($N=4$).

\subsubsection{Human Feedback: Trustworthy but Unavailable}
Across all participants, human feedback was considered the most trustworthy form of assessment. However, our participants mentioned that a real person is not always available to provide makeup feedback, and it can be difficult and cumbersome between makeup steps, so they mostly obtain human feedback after they complete the whole makeup. Our participants asked friends, family members, makeup artists, or partners to check their makeup during or after the routine. P9 explained, \textit{“After every step, I ask my sister—‘is this truly enough?’”} Others like P13 and P14 regularly used FaceTime to send images to a trusted person for review, especially when trying new colors or attending important events. P14 shared, \textit{“I ask AI too, but I always send it to a friend first.”} Still, relying on others came with logistical barriers. Specifically, our participants mentioned that they prefer to ask people who know them better regarding personalized makeup routines and preferences and can provide customized feedback. P4 said: \textit{“I prefer to go to the same agent in the Chanel counter, they know I dislike talking about aging-related products and my color preferences without having me explain everything again and again.”}

\begin{figure}[t]
    \centering
    \includegraphics[width=0.5\columnwidth]{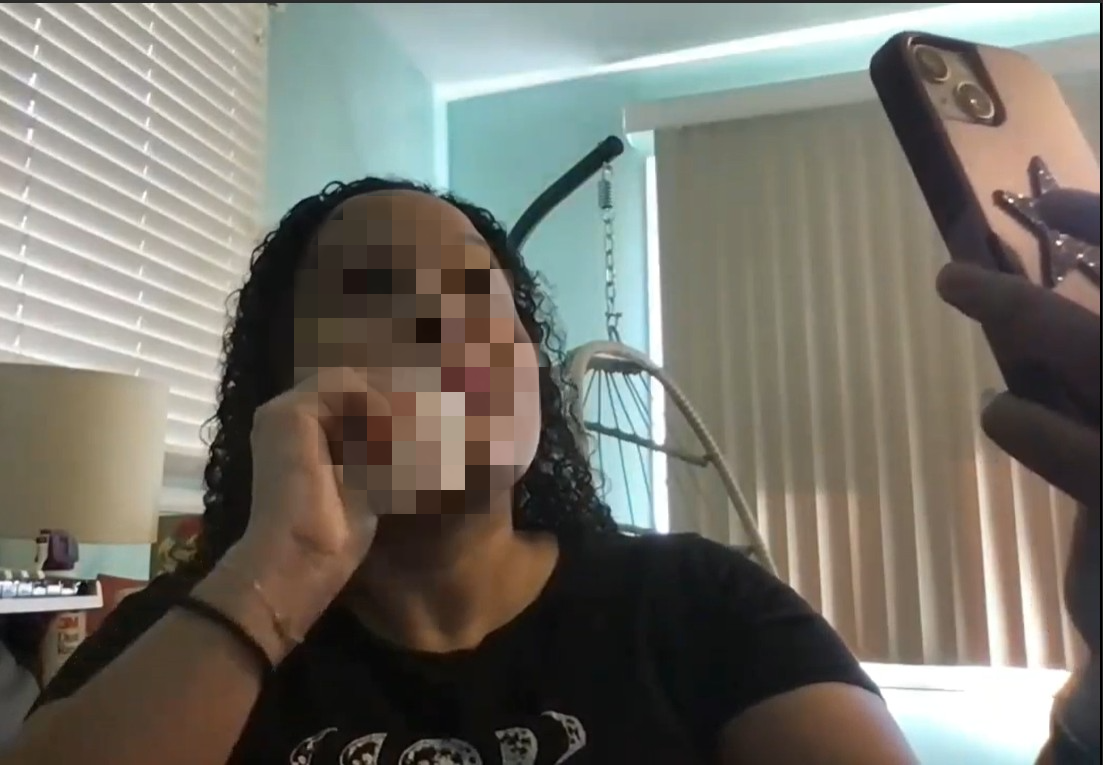}
    \caption{P7 is using EL-VMA app \cite{EstéeLau38:online} from her phone to obtain feedback on eye shadow.}
    \label{fig:assessment}
    \Description{A person with curly hair is sitting indoors, holding a smartphone with a star-shaped phone grip in one hand and applying makeup with a sponge or applicator to their cheek with the other hand. The background shows a hanging chair, vertical blinds, and a window with horizontal blinds. The person's face is pixelated for privacy.}
\end{figure}

\subsubsection{AI-Based Assessment Tools: Promising but Incomplete}
Our participants used a range of AI tools, including Seeing AI \cite{SeeingAI35:online},  BeMyAI \cite{Introduc5:online}, and Estée Lauder Voiced-enabled Makeup Assistant (EL-VMA) (Figure \ref{fig:assessment}) \cite{EstéeLau38:online}, to evaluate makeup—but with limitations in makeup applications. P7, a regular user of these apps, explained that you must prompt very specifically: \textit{“BeMyAI said there was no makeup. You really have to prompt.”} Even when prompted, participants felt the feedback was often generic or inaccurate. P12 shared, \textit{“I took a photo and asked Estée Lauder vision app—it said ‘looks great.’ Then I used BeMyAI and it found a spot on my forehead.”} She noted, however, that neither caught her eye makeup error. This inconsistency led our participants to treat AI feedback as partial, not definitive: \textit{“I can rely on it, but I cannot rely on it completely,”} said P7. After finding the responses being useless, our participants often directly abandon the AI-based assessment tool immediately without trying more. P12 said: \textit{“I will stop using the AI once I figured out it is not giving me the right answer.”}

Moreover, AI often lacked the specificity users needed. P14 pointed out that current tools can recognize basic makeup elements like lipstick but cannot answer complex aesthetic questions: \textit{“Does my contour look muddy? Is it blended well? Where should I blend?”} Similarly, P11 expressed a desire for contextual feedback during makeup: \textit{“If there’s an app, I would like to know during makeup, not after.”}

\subsubsection{Tactile Self-Assessment: Available but Low Resolution}
In the absence of trusted humans or usable tech, our participants leaned on internal methods—especially touch-based feedback and routine familiarity. P15 described how she relies on the feel of her brush and the motion she’s used to: \textit{“I know on brushes how deep or how light to go.”} Similarly, P13 guided her brush placement with her hands, noting, \textit{“I feel the cheekbone first, then apply.”} However, blending and visual gradients posed problems. Our participants mentioned that tactile feedback can’t help with determining symmetry, matching sides, or evaluating whether one eye has more product than the other.

\subsubsection{Navigating Between Assessment Methods: Workarounds, Frustrations, and Gaps}
Overall, our participants combined all available strategies—human, tech, and touch—but still faced major limitations in accuracy, access, and efficiency. For many, assessment became a layered and time-consuming process. Our participants described taking pictures, switching apps, calling others, or guessing based on touch—often within the same session. P12 shared that she might apply mascara, check with BeMyAI, call AIRA if needed, and still ask someone in person afterward. But many apps were unreliable: they misidentified colors (P8), couldn’t recognize subtle areas like under-eye blending (P12), or gave generic compliments rather than helpful critique (P13, P14).

These multi-step workarounds led to fatigue and inefficiency. P11 said, \textit{“I don’t have time to do it well with tech while doing the makeup.”} Therefore, this leads to the importance of \textbf{providing full support of makeup feedback and recommendation system with hands-free interactions}.

\subsection{Metrics of Assessment among Makeup Procedures}
\label{Metrics of Assessment among Makeup Procedures}
According to our participants, success in makeup was not defined by achieving visual perfection but by their ability to navigate the routine confidently, assess each step with clarity, and feel prepared for everyday or social interactions. We found that they evaluated their work in three distinct phases: during individual steps, after individual steps, and after the full routine.

\subsubsection{During Individual Steps: Real-Time Correction and Control}
At the beginning of the routine, our participants wanted to stay focused and in control of the process. Success meant progressing smoothly through the steps, confidently using the right products, and catching mistakes before they disrupted the routine. Interruptions—like losing track of products or second-guessing progress—broke that flow and increased cognitive load. The following needs and strategies emerged as central to supporting participants during individual steps of their routine:

\begin{itemize} \item \textbf{Enable uninterrupted flow and minimal distraction during application.} Participants preferred to follow their rhythm without stopping for constant input. P7 said, \textit{“Sometimes I listen to music while doing makeup,”} highlighting the importance of ambient or low-interference feedback that fits naturally into their routine.

\item \textbf{Support real-time recognition and correction of mistakes.} Detecting errors early helped participants preserve the overall look and their sense of confidence. As P9 put it, \textit{“During the makeup, you can directly correct it,”} emphasizing that real-time insight prevents small mistakes from becoming harder to fix later.

\item \textbf{Maintain clear orientation of tools and steps in the routine.} Losing track of what product had been used or where they were in the process disrupted flow. P8 shared, \textit{“I don’t want to go back and find which one is the eyeliner. I want to know during.”} Success included staying mentally and physically aligned with each product and step.

\item \textbf{Allow for help to be accessible without requiring dependency.} While autonomy was valued, participants still wanted the option to access help on demand. P4 expressed this balance clearly: \textit{“Support should be always available because I might need help at any time.”} \end{itemize}

\subsubsection{After Individual Steps: Enabling Step-Level Evaluation and Adjustment}
Once a product was applied, our participants paused to verify that it had been placed correctly, blended evenly, and matched across both sides of the face. Success in this stage meant avoiding subtle errors that could go unnoticed but impact the final look. It also meant knowing when to move forward with confidence. In this phase, participants prioritized the following types of feedback and support:

\begin{itemize} \item \textbf{Check for symmetry and alignment across facial features.} When applying products like blush or eyeshadow, participants wanted to ensure both sides looked consistent. P1 emphasized, \textit{“I want symmetrical analysis and feedback,”} showing that this visual concept was central even in non-visual routines.

\item \textbf{Confirm product placement and blending quality.} Participants relied on touch, but noted that blending errors could still slip through. P2 said, \textit{“I check with my fingers to assess if the makeup is in the right place,”} but blending remained one of the hardest qualities to assess tactilely.

\item \textbf{Identify specific areas that require correction.} Generic feedback—like “looks fine”—was not helpful. Participants wanted to know exactly what to fix and where. P3 noted, \textit{“Spot where there is off,”} and P6 added, \textit{“Identify spots where not much color were off.”}

\item \textbf{Provide feedback that aligns with user intent and detail level.} Participants expected feedback that could match their desired style and intensity. P7, after receiving vague input, said, \textit{“Maybe I didn’t put enough eyeshadow on, but the app said it looked normal.”} Successful step feedback should be nuanced and grounded in the user’s intent. \end{itemize}

\subsubsection{After the Full Routine: Ensuring Closure and Confidence}
After completing their makeup, our participants wanted a clear sense of closure: confirmation that they had applied everything they intended, that the look was suitable for their day, and that nothing was glaringly off. The goal was not perfection, but confidence, peace of mind, and readiness. To achieve closure and confidence after completing their makeup, participants described the need for:

\begin{itemize} \item \textbf{Indicate when the overall look is complete.} Several participants expressed uncertainty about whether their routine was truly done. P5 shared, \textit{“There’s no way to confirm how good it looks like,”} pointing to the need for a clear sense of completeness or reassurance before leaving the mirror.

\item \textbf{Support both independent self-check and collaborative validation.} Our participants alternated between self-reliance and asking others, depending on the occasion and their own experience. P5 said, \textit{“I’ll get immediate confirmation if I did correctly or not,”} while P6 noted, \textit{“I take pictures and use a magnifier to see the result.”}

\begin{table*}[t]
\centering
\resizebox{\textwidth}{!}{%
\begin{tabular}{>{\raggedright\arraybackslash}p{5cm} >{\raggedright\arraybackslash}p{6cm} >{\raggedright\arraybackslash}p{6.5cm}}
\toprule
\textbf{Category} & \textbf{Description} & \textbf{Example} \\
\midrule
Step Confirmation and General Feedback &
Confirmation of whether a step was completed correctly or achieved the intended effect. &
\textit{“Am I completing this step?”} (P1) \newline \textit{“Is it applied properly?”} (P5) \\

Color and Shade &
Identification, coordination, and appropriateness of color choices. &
\textit{“Am I using the right colors for my skin?”} (P6, P13) \newline \textit{“Does this blush look too dark?”} (P12) \\

Placement and Location &
Assistance with determining correct product placement on the face. &
\textit{“Is this the right position for blush?”} (P1, P4, P9) \newline \textit{“Where should I apply the highlighter?”} (P4) \\

Symmetry and Alignment &
Ensuring even application and facial symmetry. &
\textit{“Is one side darker than the other?”} (P3) \newline \textit{“Is everything applied evenly?”} (P10, P13, P14, P15) \\

Quantity and Amount &
Feedback on how much product to apply or whether it’s too much. &
\textit{“How do I know how much is on the brush?”} (P12) \newline \textit{“What is the right amount to apply?”} (P9, P11) \\

Blending and Integration &
Concerns about smoothness of blending and integration between colors. &
\textit{“Is it blended properly with my other makeup?”} (P5) \newline \textit{“How do I use four colors and blend them correctly?”} (P8) \\

Instruction and Walkthroughs &
Requests for step-by-step help or detailed guidance. &
\textit{“Can you walk me through applying eyeliner?”} (P5, P13, P15) \newline \textit{“How do I apply eyeliner under the eyelash line?”} (P15) \\

Technique and Strategy Exploration &
Exploring alternative methods or improving precision. &
\textit{“Is there a better way to apply this?”} (P5, P11, P13) \newline \textit{“What’s a good technique for sharper lines or lower lashes?”} (P12) \\

Tools and Product Recommendations &
Requests for specific product or tool suggestions and alternatives. &
\textit{“What brush is best for this task?”} (P1, P7, P8, P12) \newline \textit{“Is there a concealer bottle that’s easier to use?”} (P15) \\

Mistake Detection and Hazards &
Identification of errors, product smudging, or physical discomfort. &
\textit{“Do I have mascara on my nose or cheeks?”} (P4, P12) \newline \textit{“Is lipstick on my teeth?”} (P14) \\

Adjustment and Overall Assessment &
General check-ins to assess completeness or identify fixes. &
\textit{“Can you check if anything needs adjustment?”} (P5) \newline \textit{“What product needs fixing, and what action should I take?”} (P3) \\
\bottomrule
\end{tabular}%
}
\caption{Taxonomy of Questions and Feedback Needs for Non-visual Makeup. This table categorizes the types of questions participants asked along with descriptions of the question type and related examples.}
\label{tab:taxonomy}
\end{table*}

\item \textbf{Provide a verbal or auditory summary of the completed look.} A final summary helped participants confirm alignment with their intention and style. P7 described this as part of her routine: \textit{“I use BeMyEyes, take a picture and close my eyes to listen if it describes my eyes.”}

\item \textbf{Deliver concise, actionable final feedback.} At this stage, participants didn’t want long-winded evaluations. They wanted quick cues for whether to fix something or proceed. As P5 put it, \textit{“I don’t want it long. Just short, actionable feedback.”}

\item \textbf{Balance honesty with emotional reassurance.} Participants valued directness, especially when it helped avoid embarrassment later. P6 said, \textit{“If I’m doing it wrong, I need someone to tell me. I’m not gonna dissolve.”} Feedback needed to be respectful, but not sugar-coated.

\item \textbf{Account for contextual factors like lighting changes.} Makeup that looked fine in one environment could fail in another. As P6 explained, \textit{“Light is difficult—if you go to a restaurant that is dark.”} Supporting success meant acknowledging these situational shifts. \end{itemize}

\renewcommand{\arraystretch}{1.3}

\subsection{Taxonomy of Questions and Feedback Needs for Non-visual Makeup}
\label{Taxonomy of Questions and Feedback Needs for Non-visual Makeup}
\begin{figure}[t]
    \centering
    \includegraphics[width=1\columnwidth]{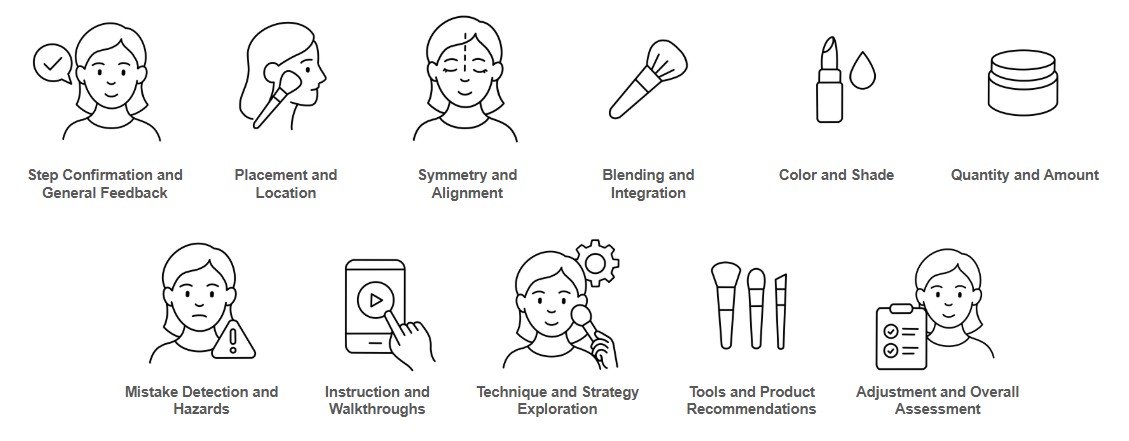}
    \caption{Categories of questions and feedback needs during non-visual makeup routines, as defined in our taxonomy (see Table 3). This visual highlights the diversity of information participants sought.}
    \label{fig:illustrationofquestions}
    \Description{A set of eleven black-and-white icons representing different types of questions that people with vision impairments ask during makeup routines. The icons correspond to: (1) Step Confirmation and General Feedback — a face with a checkmark bubble; (2) Placement and Location — a person applying blush on their cheek; (3) Symmetry and Alignment — a face with alignment marks; (4) Blending and Integration — a makeup brush; (5) Color and Shade — a lipstick and color droplet; (6) Quantity and Amount — a cream jar; (7) Mistake Detection and Hazards — a worried face with an exclamation mark; (8) Instruction and Walkthroughs — a hand pointing to a video play button; (9) Technique and Strategy Exploration — a person experimenting with makeup on their face; (10) Tools and Product Recommendations — three makeup brushes; (11) Adjustment and Overall Assessment — a smiling face with a checklist.

}
\end{figure}

\begin{figure}[t]
    \centering
    \includegraphics[width=0.8\columnwidth]{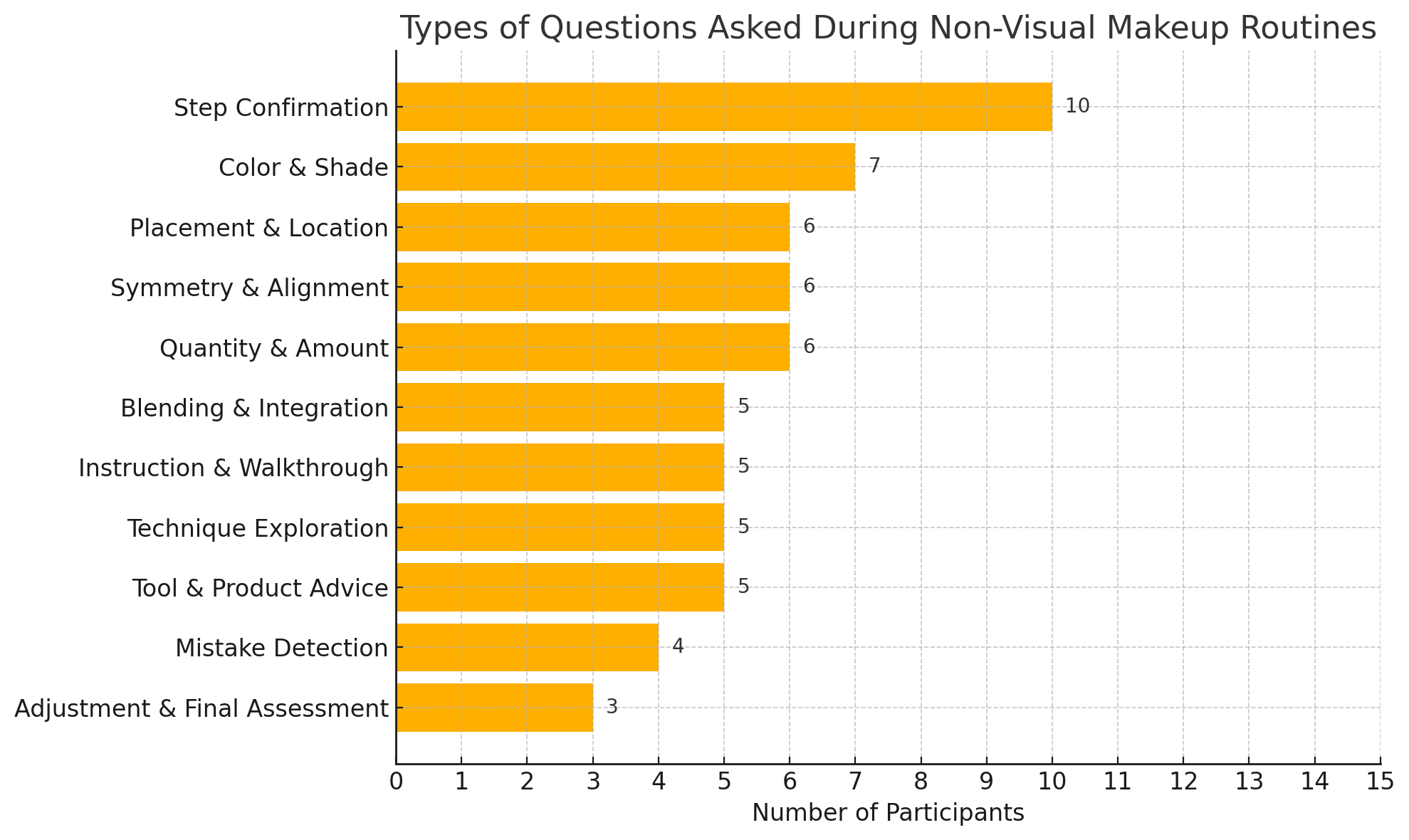}
    \caption{Frequency of each question type asked by participants during non-visual makeup routines. This figure emphasizes which feedback needs were most common in our study.}
    \label{fig:questiontype}
    \Description{Bar chart showing the number of participants (out of 15) who asked various types of questions during non-visual makeup routines. The most common question type was "Step Confirmation" (10 participants), followed by "Color \& Shade," "Placement \& Location," "Symmetry \& Alignment," and "Quantity \& Amount" (each with 6 or 7 participants). Other categories include "Blending \& Integration," "Instruction \& Walkthrough," "Technique Exploration," "Tool \& Product Advice" (5 participants each), "Mistake Detection" (4), and "Adjustment \& Final Assessment" (3). The chart highlights the wide range of informational needs during makeup application.}
\end{figure}

Throughout their makeup routines, participants frequently posed questions to confirm their progress, evaluate their appearance, and refine their techniques. These inquiries surfaced across all stages of application and reflect the types of support they actively sought—ranging from moment-to-moment confirmation and correction to higher-level adjustments and strategic decision-making (Figure~\ref{fig:illustrationofquestions}). By grouping and analyzing these questions, we derive a concrete framework for understanding the informational demands that assistive systems must be prepared to meet—whether by providing real-time feedback, step-by-step instruction, or post-application assessment (Table \ref{tab:taxonomy}). As shown in Figure~\ref{fig:questiontype}, the most commonly asked questions related to step confirmation ($N=10$), followed by concerns around color selection ($N=7$), product placement ($N=6$), symmetry ($N=6$), and quantity ($N=6$). This distribution underscores the importance of addressing both procedural guidance and aesthetic evaluation throughout the makeup process. The taxonomy introduced here is grounded in participants’ actual language and lived experience, not hypothetical scenarios (Table \ref{tab:taxonomy}).

This taxonomy of questions (Table~\ref{tab:taxonomy}) reveals the depth and breadth of feedback and support needed in non-visual makeup routines. These inquiries mark critical junctures of decision-making, self-correction, learning, and creative expression. They span both practical challenges—such as achieving even symmetry, applying the correct quantity, or selecting suitable colors—and more strategic goals like refining techniques, seeking instruction, and assessing the final look (Table~\ref{tab:taxonomy}). Grounded in real-world user behavior, this taxonomy highlights actionable opportunities for assistive technologies to deliver timely, relevant, and context-sensitive support. Any system designed to assist with non-visual makeup should be flexible enough to accommodate this diverse range of informational needs.

Moreover, the taxonomy highlights that non-visual makeup is not merely about converting visual instructions into audio cues. It demands multilayered feedback mechanisms that mirror the tactile, spatial, and aesthetic complexities of the process. Questions about blending, placement, and symmetry expose the fine-grained coordination required across different facial areas, while concerns about color and product amount speak to a need for aesthetic calibration without visual reference (Table~\ref{tab:taxonomy}). These patterns suggest that effective support must move beyond binary validation (e.g., “yes” or “no”) to deliver nuanced, user-aligned feedback that reflects intent, context, and personal style.

Finally, our participants frequently expressed a desire not just for guidance, but for opportunities to learn, experiment, and grow (Table~\ref{tab:taxonomy}). Categories such as technique exploration, product recommendations, and final assessment point toward a mindset of iterative improvement and intentional decision-making. Users weren’t just trying to get makeup “right”—they were developing their own strategies, routines, and aesthetic preferences. Assistive systems, then, should not be limited to instructional or corrective roles, but instead adopt a collaborative stance—acting as responsive partners that evolve with the user, offering feedback, encouragement, and adaptation as needs and goals change over time.

\subsection{Expert Reflections: Affirming Embodied Strategies and Exploring Accessible Alternatives}
\label{Expert Reflections: Affirming Embodied Strategies and Exploring Accessible Alternatives}
While participants in our study demonstrated rich, embodied expertise in navigating non-visual makeup routines, we engaged professional makeup artists to reflect on these practices and offer additional perspectives in a collaborative, non-hierarchical manner. Importantly, we position their feedback not as prescriptive instruction or correction, but as a means to affirm participants' strategies, validate tactile-first expertise, and collaboratively explore accessible alternatives that can augment—rather than diminish—user agency. Our approach emphasizes mutual respect and positions professional insights as resources that support user-defined goals and preferences.

\subsubsection{Tactile-First Application is Already Best Practice}

Makeup artists consistently affirmed the value of touch and muscle memory as integral components of participants’ routines. Rather than viewing finger-based application as a workaround for the absence of vision, they highlighted it as a deliberate and effective technique used by professionals. M1 remarked, \textit{“Using your hands is the best way to warm up the product, and it gives you immediate feedback on where it’s going.”} Similarly, M2 emphasized that tactile strategies are not unique to non-visual routines: \textit{“Even sighted users benefit from touch—muscle memory and tactile feedback are how we learn makeup.”} One participant, P3, was specifically commended for her tapping motion during foundation application, which M2 described as \textit{“already a great professional trick.”} These expert perspectives validate tactile-first approaches not as compensatory, but as skillful and intentional methods that can produce precise and polished results.

\subsubsection{Product Textures That Support Sensory Control}

Makeup artists consistently recommended cream and liquid-based products as more accessible and intuitive for non-visual application than powder formulations. These textures offer greater tactile feedback, are easier to control, and allow for gradual, buildable application. M3 noted, \textit{“Creams are easier to feel and blend—less fallout, less pressure required.”} M1 emphasized the functional advantage of cream sticks: \textit{“Cream sticks, especially for eyeshadow or contour, let you use your fingers without needing a separate tool.”} Similarly, M4 described the sensory familiarity of cream textures, stating, \textit{“Paint pot cream shadows feel like lip balm—they’re forgiving and easy to manipulate.”} These insights highlight the importance of selecting products that align with tactile engagement and physical control, rather than relying solely on visual aesthetics. Such recommendations can significantly reduce complexity while supporting precision and independence in makeup routines.

\subsubsection{Tool Control and Ergonomics Outweigh Tool Type}

Rather than advocating for a specific tool—such as a brush, sponge, or finger—makeup artists emphasized that control, grip, and pressure sensitivity are the most critical factors in successful application. M4 explained, \textit{“A smaller brush is easier to control—big brushes feel disconnected,”} underscoring how tool size and ergonomics can significantly impact user precision. Similarly, M5 suggested that \textit{“she would benefit from mini brushes or lip crayons—those give more control.”} These insights echo participants’ experiences, reinforcing that well-designed, intuitively sized tools can enhance accuracy and reduce reliance on visual feedback. Optimizing tool ergonomics can therefore play a key role in supporting independent and confident makeup routines.

\subsubsection{Error Prevention Through Procedural Sequencing}

Makeup artists emphasized that while sighted users can rely on visual inspection to correct mistakes, blind users benefit significantly from routines that are strategically ordered to prevent errors before they occur. M4 noted, \textit{“This is a standard in professional routines, but even more critical when you can’t easily go back to fix a mistake,”} referencing practices like applying eyeshadow before foundation to avoid fallout. M5 reinforced this point, stating, \textit{“Doing all cream steps before powder is essential—it prevents caking and makes everything easier to blend.”} These observations underscore the importance of not only accessible product guidance but also clear, intentional sequencing. Assistive systems should therefore support users by conveying not just how to apply makeup, but when and in what order, enabling more reliable, efficient, and error-resistant routines.

\subsubsection{Challenges in Symmetry and Blending}

Among all aspects of non-visual makeup application, blending and achieving facial symmetry emerged as the most consistently challenging tasks—particularly for products like blush, contour, and eyeliner. These steps require subtle, spatially distributed judgment that is difficult to assess through touch alone. M1 noted, \textit{“Blending is the hardest for blind users—you can't rely on bone structure the same way.”} M2 stressed the importance of developing structured techniques over time: \textit{“It takes time to develop a tactile sense of symmetry—you need a formula for the face.”} M3 echoed this by emphasizing the need for tactile sensitivity, explaining, \textit{“Users have to learn to feel for areas where blending is uneven—this is where flashback or harsh lines can sneak in.”} These observations underscore the need for assistive tools that provide feedback on symmetry and blending, as well as instructional supports that help users develop repeatable, tactile-based strategies for achieving even, polished results.

\subsubsection{Strong Techniques but with Gaps in Targeted Product Knowledge}

Makeup artists consistently praised participants for their intuitive techniques, adaptive strategies, and self-awareness in navigating non-visual makeup routines. However, they also identified a recurring gap in targeted product knowledge—particularly around selecting products and tools that align with their individual needs. M1 observed from P5 makeup routine and commented, \textit{“She clearly understands the routine, but some of her products weren’t the best fit—matching foundation shades or picking a blush formula that’s easier to control would really help.”} Similarly, M2 noted, \textit{“They’re very resourceful, but would benefit from more education on what each brush or formula is designed to do.”} This feedback highlights the importance of accessible resources that go beyond application techniques, offering guidance on how to evaluate, compare, and choose products that optimize control, ease of use, and overall effectiveness for non-visual application.

\subsubsection{Accessible Alternatives Exist but Lack Visibility and Framing}

Makeup artists identified numerous tools and techniques already used in professional practice that could greatly benefit blind users—yet are rarely introduced or promoted within accessible contexts. M3 remarked, \textit{“Some of the best tools are already out there, but blind users don’t know they exist because no one frames them that way.”} Recommendations included items such as mascara shields to prevent smudging, angled brushes for more precise contouring, and travel-sized products that enhance grip and control. M4 pointed out that \textit{“Real Techniques brush sets are affordable and labeled by use—they’re great for tactile exploration.”} These insights reveal a critical opportunity: to identify and reframe existing, effective tools in ways that align with the priorities and needs of blind users. By curating and promoting such products with accessible guidance, designers and educators can bridge the gap between mainstream beauty tools and inclusive, non-visual makeup routines.


\subsection{Preferred Communication and Interaction for Makeup Assessment and Feedback}
\label{Preferred Communication and Interaction for Makeup Assessment and Feedback}

While the content of feedback was critical, our participants also emphasized how they wanted that information delivered. Their preferences for communication methods reflected needs for accessibility, independence, fluidity, and emotional support.

\subsubsection{Conversational and Voice-Based Interfaces}

All participants mentioned using a voice-based or conversational assistant to help guide their makeup process. Rather than relying solely on visual interfaces, participants preferred interacting in natural language, with the option to ask follow-up questions or request clarification. P3 and P4 favored conversational agents, and P5 offered a detailed imagined use case: \textit{“I would say ‘App name, record,’ and once I’m done, then it gives me feedback on how I applied the blush… Then I can ask more questions like ‘does it interfere with my concealer?’”} The appeal was not just convenience, but also the ability to customize the interaction flow.

\subsubsection{Support During the Process, Not Just Afterward}

Real-time guidance was highly valued. Many participants wanted the ability to check on their application as they went, reducing the risk of needing to redo steps. P6 emphasized, \textit{“I don’t want someone to wait until the end.”} Instead, she preferred suggestions \textit{“while I’m doing the makeup—ways to make it better or easier.”} P14 added that she often finds herself checking in visually mid-process: \textit{“I can show my face and quickly make adjustments or corrections.”}

\subsubsection{Control and Autonomy in Interaction Style}

Participants varied in how much control they wanted over system behavior. Some preferred passive tools that offered feedback only when asked, while others were more open to continuous monitoring. P7 explained, \textit{“I like Estée Lauder’s app—it’s more passive. I don’t want instruction every time.”} However, she also noted, \textit{“I don’t interact while doing it—I’m confident while doing it.”} This flexibility was key—participants wanted to maintain a sense of agency and confidence.

\subsubsection{Environment and Mood-Aware Interaction}

Participants highlighted the importance of considering the broader environment during interaction. P8 noted that \textit{“lighting matters—make sure you’re in the right lighting to do color matching clearly.”} Others wanted the tone of interaction to match their emotional state or energy level. P11 appreciated how ChatGPT’s style sometimes aligned with hers: \textit{“The conversation should match my vibe—calm but positive.”}

\subsection{Summary}
Our participants’ reflections reveal a clear vision for what an ideal non-visual makeup support system should provide: personalized, honest, and context-aware feedback delivered through flexible, hands-free interaction. Beyond basic instruction, they emphasized the value of guidance that aligns with their identity, goals, and routines—reinforcing autonomy while offering expert-level insight. These preferences point to the need for systems that are not only technically capable but emotionally attuned and situationally responsive.

\section{Discussion}
In the discussion, we reflect on the broader implications of our findings — redefining accessible beauty, identifying system needs for procedural support, and proposing design considerations that respect user agency, learning, and identity.

\subsection{Accessible Beauty is Not About Simplicity — It’s About Strategy, Adaptation, and Embodied Expertise}
In accessibility design, simplicity is frequently treated as a proxy for usability. However, our findings challenge the assumption that reducing complexity is always the optimal path. For people with vision impairments, makeup is not about avoiding difficulty — it is about cultivating control, mastery, and personalized strategies over time. Participants in our study demonstrated rich embodied expertise, developing tactile-first techniques like finger-based application, blending by touch, and memorized routines to navigate non-visual makeup. These practices align with prior HCI work on user appropriation \cite{shinohara2011shadow} and tactile expertise \cite{li2022feels}, where people with disabilities transform constraints into opportunities for learning and self-expression.

This challenges dominant assistive technology narratives focused on compensating for "missing" sensory input. Our findings show that users do not simply need visual substitutes — they need systems that collaborate with their existing sensory ecologies across all makeup procedures interdependently \cite{bennett2018interdependence}. Assistive technologies should scaffold users' embodied knowledge, not overwrite it. Accessible beauty, then, is not about removing complexity but supporting strategic adaptation and amplifying user agency. Future systems should move beyond deficit-compensation models toward tools that respect and enhance users' procedural expertise, enabling creativity, independence, and self-expression on their own terms.

\subsection{Procedural Feedback Systems as a New Interaction Paradigm}
Our study positions procedural feedback systems as a distinct and underexplored class of assistive technology — moving beyond task-based support toward process-oriented guidance. While existing accessibility tools excel at helping users identify objects, read labels, or detect colors \cite{mascetti2016towards, li2022feels}, these solutions operate at isolated moments of need. They do not address the broader procedural flow of complex tasks like makeup application, where users must navigate temporal sequencing, spatial management, sensory calibration, and real-time decision-making.

Non-visual makeup routines are inherently procedural — requiring users to coordinate multiple factors simultaneously, such as tracking product order, maintaining spatial awareness of the face, and adjusting pressure and technique through touch. Participants described the cognitive load of managing these steps without visual feedback, underscoring the need for systems that can support them throughout the process, not just in isolated moments. We argue that future assistive systems should act as intelligent co-pilots — reasoning about temporality, step dependencies, error recovery pathways, and user-defined goals. Procedural feedback systems must monitor user progress, anticipate challenges, and provide actionable, context-sensitive guidance. Designing such systems requires moving beyond static interactions toward dynamic, adaptive scaffolding — leveraging models of procedural flow, user history, and real-time sensory data to enable users to manage complex tasks independently and confidently.

\subsection{Redefining Success and Precision in Non-Visual Makeup}
Our findings challenge conventional metrics of success in makeup application — often rooted in visual-centric ideals such as perfect symmetry, flawless coverage, or precise color matching \cite{pradhan2021inclusive, kennedy2016exploring}. For participants in our study, success was rarely about achieving an objective visual standard; rather, it was deeply personal, contextual, and relational. A successful makeup outcome was defined not by uniformity or precision alone, but by how well the final appearance aligned with their individual goals, comfort, and self-expression within a given context.

Existing AI-based feedback systems for makeup—where available—typically offer generic or objective validation (e.g., confirming whether foundation is applied or a step is completed). However, our participants often found such feedback unhelpful or irrelevant to their actual concerns. For example, P12 described AI feedback as “useless” because it failed to address questions about specific visual qualities such as blending or evenness. Similarly, P14 expressed the need for feedback on nuanced aesthetic issues, such as asking, “Does my contour look muddy?” These examples illustrate the gap between current AI capabilities and the real, nuanced needs of users. Bridging this gap presents a substantial challenge for AI: moving from simple task verification to providing context-aware, aesthetic feedback on complex visual attributes like gradients, blending, or evenness. This requires not only advanced visual analysis but also a deeper alignment with user-defined goals and subjective standards. Addressing this challenge is crucial for developing assistive technologies that genuinely empower users in their makeup routines.

This reframing calls for new evaluation frameworks in assistive beauty technologies — frameworks that foreground subjective experiences, situational appropriateness, and user-defined aesthetics over rigid aesthetic ideals. For instance, the "perfect" makeup look for a casual day at home may prioritize speed, ease, or emotional grounding, while makeup for a formal event may emphasize confidence or intentional expressiveness. Future systems should move toward participatory and adaptive metrics of success — allowing users to rate their satisfaction, confidence, or preparedness with their final look. Rather than positioning the system as an external evaluator of correctness, feedback systems should act as collaborative partners that help users reflect on whether their appearance aligns with their intended outcome. In doing so, designers can create technologies that not only accommodate diversity in beauty practices but actively celebrate and support it — empowering users to define success on their own terms, across different moods, identities, and social situations.

\subsection{Trust, Honesty, and Emotional Framing Are Central to Feedback Delivery}

Beyond technical accuracy, the emotional tone of feedback emerged as a critical dimension for trust, engagement, and sustained system use \cite{li2022feels}. Participants consistently expressed frustration with feedback that was vague, overly generic, or patronizing. Feedback that simply offered empty praise ("Looks good!") without actionable content was perceived as dismissive or untrustworthy. In contrast, participants valued feedback that was direct, constructive, and respectful — offering honest evaluations while acknowledging their effort and agency.

This highlights the need for feedback systems that are not only technically capable but emotionally intelligent. Effective systems should balance correction with affirmation, providing clear guidance while celebrating users' strategies, adaptations, and progress. Mistakes should be framed not as failures but as natural parts of learning and self-exploration. Moreover, emotional framing is not one-size-fits-all. Future systems should offer customization of feedback styles — allowing users to select their preferred tone or mode of interaction (e.g., gentle encouragement, direct coaching, or neutral instruction). This personalization respects the diverse emotional needs, learning styles, and aesthetic values of users, fostering trust and long-term engagement in beauty routines.

\subsection{Hands-Free, Conversational Interaction Enhances Independence}

Non-visual makeup routines exemplify embodied, tactile-first tasks where hands-free, voice-based interaction is not merely convenient — it is foundational for accessible system design. Future Procedural Feedback Systems must support fluid, conversational interaction modalities that align with users' physical constraints, multi-hand operations, and real-time focus \cite{truong2021automatic,chang2021rubyslippers}. Our participants envisioned rich conversational models where they could engage in flexible, on-demand interaction with the system — initiating feedback, pausing instructions, querying for specific details, or requesting post-step evaluations while maintaining focus on the application itself. Crucially, these interactions should not impose rigid structures or interruptive prompts but adapt to the user's pace, routine, and preferences. Future systems should support a range of interaction styles — from continuous guidance for novice users, to feedback-on-demand for experienced routines, to periodic check-ins for specific steps. Natural speech interfaces, multi-turn dialogue systems, and context-aware voice controls can play a vital role in enhancing user autonomy and reducing reliance on sighted assistance.

\subsection{Interaction Design Principles for Supporting Procedural Practices}
Building on the need for process-oriented guidance, we outline key interaction design principles for assistive systems that support non-visual makeup as a procedural, embodied practice. These principles emphasize how systems should behave during interaction — enabling flexible engagement, adaptive support, and real-time scaffolding that aligns with users' routines and expertise.

\subsubsection{Support Fluid Transitions Across Steps}
Non-visual makeup routines involve managing temporal sequences, spatial navigation, and task switching without visual anchors. Systems should help users track progress, maintain orientation, and transition smoothly between steps. This might include lightweight reminders of the current step ("You're applying blush now"), optional previews of upcoming steps, or gentle cues for missing steps.

\subsubsection{Enable Granular, On-Demand Guidance}
Rather than enforcing rigid instruction flows, systems should support layered guidance — allowing users to request feedback at varying levels of specificity. Participants valued control over how much help they received — from high-level validation to detailed, step-by-step breakdowns. Interaction models should allow users to query for clarification without disrupting their routine.

\subsubsection{Respond to User-Specific Routines and Techniques}
Users developed highly personalized, embodied routines — from tactile blending techniques to idiosyncratic product orders. Systems should avoid assuming standardized sequences or techniques. Instead, they should learn from user history and adapt feedback to reinforce existing practices while supporting exploration of new techniques when requested.

\subsubsection{Balance Real-Time Assistance with Cognitive Flow}
Given the attentional demands of makeup, systems should minimize interruptions and respect users' cognitive flow. Feedback should be concise, actionable, and delivered at appropriate breakpoints — such as after completing a step or during natural pauses. Real-time assistance should enhance, not fragment, users' embodied focus.

\subsubsection{Position the System as a Collaborative Partner}
Finally, systems should position themselves not as evaluators but as collaborative partners — scaffolding user expertise rather than overriding it. This relational framing aligns with prior work on cooperative AI and human-centered automation, where technology augments human agency rather than controlling behavior. Systems should respect user-defined goals, invite negotiation, and foster a sense of partnership in the beauty process. Beyond partnering with individual users, systems should also support collaboration with trusted others—such as friends or family—by enabling seamless sharing of specific moments or questions for human feedback. This design approach recognizes the ongoing value of social input and frames assistive technology as a facilitator of both independent and socially-supported routines.

\section{Limitation}
While this study offers in-depth insights into the procedural behaviors and support needs of visually impaired makeup users, several limitations should be acknowledged. The participant sample was composed of individuals with existing interests and experience in cosmetics, which may exclude perspectives from those who have disengaged from makeup due to accessibility barriers. Additionally, all participants identified as female, limiting the applicability of findings to individuals of other gender identities. Participants in this study also varied in the onset and duration of vision impairment (see Table 1), which may influence learning pathways, adaptive strategies, and the extent to which prior visual memory shapes non-visual makeup practices. Future research could more systematically examine these differences to better understand their impact on support needs and technology design. Moreover, makeup practices and aesthetic preferences are deeply shaped by cultural, regional, and social contexts — including variations in color preferences, product availability, and normative beauty standards. Our study primarily reflects participants based in North America, and future work should explore how non-visual makeup practices, support needs, and system designs might differ across global and culturally diverse settings.

\section{Conclusion}
This work presents a contextual inquiry into the real-world makeup practices of people with vision impairments, highlighting the procedural, sensory, and evaluative strategies they employ to navigate a task that is inherently visual. By closely observing participants’ application processes and engaging professional makeup artists in reflective feedback, we uncovered the nuanced adaptations, support needs, and design opportunities across the makeup workflow. Our findings underscore the importance of moving beyond discrete, task-specific assistive tools toward systems that support the full procedural arc of makeup application. Such systems must offer context-aware guidance, flexible interaction models, and feedback aligned with users’ personal goals and definitions of success. Ultimately, this work contributes foundational knowledge for the design of inclusive beauty technologies that respect the agency and diverse aesthetic values of visually impaired individuals.

\bibliographystyle{ACM-Reference-Format}
\bibliography{main}


\begin{thebibliography}{59}


\ifx \showCODEN    \undefined \def \showCODEN     #1{\unskip}     \fi
\ifx \showISBNx    \undefined \def \showISBNx     #1{\unskip}     \fi
\ifx \showISBNxiii \undefined \def \showISBNxiii  #1{\unskip}     \fi
\ifx \showISSN     \undefined \def \showISSN      #1{\unskip}     \fi
\ifx \showLCCN     \undefined \def \showLCCN      #1{\unskip}     \fi
\ifx \shownote     \undefined \def \shownote      #1{#1}          \fi
\ifx \showarticletitle \undefined \def \showarticletitle #1{#1}   \fi
\ifx \showURL      \undefined \def \showURL       {\relax}        \fi
\providecommand\bibfield[2]{#2}
\providecommand\bibinfo[2]{#2}
\providecommand\natexlab[1]{#1}
\providecommand\showeprint[2][]{arXiv:#2}

\bibitem[Ahmetovic et~al\mbox{.}(2016)]%
        {ahmetovic2016navcog}
\bibfield{author}{\bibinfo{person}{Dragan Ahmetovic}, \bibinfo{person}{Cole Gleason}, \bibinfo{person}{Chengxiong Ruan}, \bibinfo{person}{Kris Kitani}, \bibinfo{person}{Hironobu Takagi}, {and} \bibinfo{person}{Chieko Asakawa}.} \bibinfo{year}{2016}\natexlab{}.
\newblock \showarticletitle{NavCog: a navigational cognitive assistant for the blind}. In \bibinfo{booktitle}{\emph{Proceedings of the 18th International Conference on Human-Computer Interaction with Mobile Devices and Services}}. \bibinfo{pages}{90--99}.
\newblock


\bibitem[Ahmetovic et~al\mbox{.}(2020)]%
        {ahmetovic2020recog}
\bibfield{author}{\bibinfo{person}{Dragan Ahmetovic}, \bibinfo{person}{Daisuke Sato}, \bibinfo{person}{Uran Oh}, \bibinfo{person}{Tatsuya Ishihara}, \bibinfo{person}{Kris Kitani}, {and} \bibinfo{person}{Chieko Asakawa}.} \bibinfo{year}{2020}\natexlab{}.
\newblock \showarticletitle{Recog: Supporting blind people in recognizing personal objects}. In \bibinfo{booktitle}{\emph{Proceedings of the 2020 CHI Conference on Human Factors in Computing Systems}}. \bibinfo{pages}{1--12}.
\newblock


\bibitem[AI({[n.\,d.]})]%
        {SeeingAI35:online}
\bibfield{author}{\bibinfo{person}{Seeing AI}.} \bibinfo{year}{[n.\,d.]}\natexlab{}.
\newblock \bibinfo{title}{Seeing AI - Talking Camera for the Blind}.
\newblock
\urldef\tempurl%
\url{https://www.seeingai.com/}
\showURL{%
\tempurl}
\newblock
\shownote{[Online; accessed 2025-04-16]}.


\bibitem[Aira(2014)]%
        {HomeAira84:online}
\bibfield{author}{\bibinfo{person}{Aira}.} \bibinfo{year}{2014}\natexlab{}.
\newblock \bibinfo{title}{Home - Aira : Aira}.
\newblock \bibinfo{howpublished}{\url{https://aira.io/}}.
\newblock
\newblock
\shownote{(Accessed on 06/14/2021)}.


\bibitem[Apple(2016)]%
        {FaceTime24:online}
\bibfield{author}{\bibinfo{person}{Apple}.} \bibinfo{year}{2016}\natexlab{}.
\newblock \bibinfo{title}{FaceTime on the App Store}.
\newblock
\urldef\tempurl%
\url{https://apps.apple.com/us/app/facetime/id1110145091}
\showURL{%
\tempurl}
\newblock
\shownote{[Online; accessed 2025-04-16]}.


\bibitem[Araviiskaia et~al\mbox{.}(2022)]%
        {araviiskaia2022recommendations}
\bibfield{author}{\bibinfo{person}{Elena Araviiskaia}, \bibinfo{person}{Anne Le~Pillouer~Prost}, \bibinfo{person}{Marita Kosmadaki}, \bibinfo{person}{Delphine Kerob}, {and} \bibinfo{person}{Elia Roo}.} \bibinfo{year}{2022}\natexlab{}.
\newblock \showarticletitle{Recommendations for the use of corrective makeup after dermatological procedures}.
\newblock \bibinfo{journal}{\emph{Journal of Cosmetic Dermatology}} \bibinfo{volume}{21}, \bibinfo{number}{4} (\bibinfo{year}{2022}).
\newblock


\bibitem[Bandukda et~al\mbox{.}(2024)]%
        {bandukda2024context}
\bibfield{author}{\bibinfo{person}{Maryam Bandukda}, \bibinfo{person}{Yichen Wang}, \bibinfo{person}{Monica Perusquia-Hernandez}, \bibinfo{person}{Franklin~Mingzhe Li}, {and} \bibinfo{person}{Catherine Holloway}.} \bibinfo{year}{2024}\natexlab{}.
\newblock \showarticletitle{Context matters: Investigating information sharing in mixed-visual ability social interactions}. In \bibinfo{booktitle}{\emph{Extended Abstracts of the CHI Conference on Human Factors in Computing Systems}}. \bibinfo{pages}{1--8}.
\newblock


\bibitem[BeMyEyes(2015)]%
        {BeMyEyes41:online}
\bibfield{author}{\bibinfo{person}{BeMyEyes}.} \bibinfo{year}{2015}\natexlab{}.
\newblock \bibinfo{title}{Be My Eyes - See the world together}.
\newblock \bibinfo{howpublished}{\url{https://www.bemyeyes.com/}}.
\newblock
\newblock
\shownote{(Accessed on 06/14/2021)}.


\bibitem[bemyeyes editor(2023)]%
        {Introduc5:online}
\bibfield{author}{\bibinfo{person}{bemyeyes editor}.} \bibinfo{year}{2023}\natexlab{}.
\newblock \bibinfo{title}{Introducing Be My AI (formerly Virtual Volunteer) for People who are Blind or Have Low Vision, Powered by OpenAI’s GPT-4 - Be My Eyes}.
\newblock
\urldef\tempurl%
\url{https://www.bemyeyes.com/news/introducing-be-my-ai-formerly-virtual-volunteer-for-people-who-are-blind-or-have-low-vision-powered-by-openais-gpt-4/}
\showURL{%
\tempurl}
\newblock
\shownote{[Online; accessed 2025-04-16]}.


\bibitem[Bennett et~al\mbox{.}(2018)]%
        {bennett2018interdependence}
\bibfield{author}{\bibinfo{person}{Cynthia~L Bennett}, \bibinfo{person}{Erin Brady}, {and} \bibinfo{person}{Stacy~M Branham}.} \bibinfo{year}{2018}\natexlab{}.
\newblock \showarticletitle{Interdependence as a frame for assistive technology research and design}. In \bibinfo{booktitle}{\emph{Proceedings of the 20th international acm sigaccess conference on computers and accessibility}}. \bibinfo{pages}{161--173}.
\newblock


\bibitem[Bigham et~al\mbox{.}(2010)]%
        {bigham2010vizwiz}
\bibfield{author}{\bibinfo{person}{Jeffrey~P Bigham}, \bibinfo{person}{Chandrika Jayant}, \bibinfo{person}{Hanjie Ji}, \bibinfo{person}{Greg Little}, \bibinfo{person}{Andrew Miller}, \bibinfo{person}{Robert~C Miller}, \bibinfo{person}{Robin Miller}, \bibinfo{person}{Aubrey Tatarowicz}, \bibinfo{person}{Brandyn White}, \bibinfo{person}{Samual White}, {et~al\mbox{.}}} \bibinfo{year}{2010}\natexlab{}.
\newblock \showarticletitle{Vizwiz: nearly real-time answers to visual questions}. In \bibinfo{booktitle}{\emph{Proceedings of the 23nd annual ACM symposium on User interface software and technology}}. \bibinfo{pages}{333--342}.
\newblock


\bibitem[Brady et~al\mbox{.}(2013)]%
        {brady2013visual}
\bibfield{author}{\bibinfo{person}{Erin Brady}, \bibinfo{person}{Meredith~Ringel Morris}, \bibinfo{person}{Yu Zhong}, \bibinfo{person}{Samuel White}, {and} \bibinfo{person}{Jeffrey~P Bigham}.} \bibinfo{year}{2013}\natexlab{}.
\newblock \showarticletitle{Visual challenges in the everyday lives of blind people}. In \bibinfo{booktitle}{\emph{Proceedings of the SIGCHI conference on human factors in computing systems}}. \bibinfo{pages}{2117--2126}.
\newblock


\bibitem[Braun and Clarke(2006)]%
        {braun2006using}
\bibfield{author}{\bibinfo{person}{Virginia Braun} {and} \bibinfo{person}{Victoria Clarke}.} \bibinfo{year}{2006}\natexlab{}.
\newblock \showarticletitle{Using thematic analysis in psychology}.
\newblock \bibinfo{journal}{\emph{Qualitative research in psychology}} \bibinfo{volume}{3}, \bibinfo{number}{2} (\bibinfo{year}{2006}), \bibinfo{pages}{77--101}.
\newblock


\bibitem[Chang et~al\mbox{.}(2021)]%
        {chang2021rubyslippers}
\bibfield{author}{\bibinfo{person}{Minsuk Chang}, \bibinfo{person}{Mina Huh}, {and} \bibinfo{person}{Juho Kim}.} \bibinfo{year}{2021}\natexlab{}.
\newblock \showarticletitle{RubySlippers: Supporting Content-based Voice Navigation for How-to Videos}. In \bibinfo{booktitle}{\emph{Proceedings of the 2021 CHI Conference on Human Factors in Computing Systems}}. \bibinfo{pages}{1--14}.
\newblock


\bibitem[Chang et~al\mbox{.}(2024)]%
        {chang2024worldscribe}
\bibfield{author}{\bibinfo{person}{Ruei-Che Chang}, \bibinfo{person}{Yuxuan Liu}, {and} \bibinfo{person}{Anhong Guo}.} \bibinfo{year}{2024}\natexlab{}.
\newblock \showarticletitle{WorldScribe: Towards Context-Aware Live Visual Descriptions}. In \bibinfo{booktitle}{\emph{Proceedings of the 37th Annual ACM Symposium on User Interface Software and Technology}}. \bibinfo{pages}{1--18}.
\newblock


\bibitem[Chong et~al\mbox{.}(2021)]%
        {chong2021exploring}
\bibfield{author}{\bibinfo{person}{Toby Chong}, \bibinfo{person}{Nolwenn Maudet}, \bibinfo{person}{Katsuki Harima}, {and} \bibinfo{person}{Takeo Igarashi}.} \bibinfo{year}{2021}\natexlab{}.
\newblock \showarticletitle{Exploring a Makeup Support System for Transgender Passing based on Automatic Gender Recognition}. In \bibinfo{booktitle}{\emph{Proceedings of the 2021 CHI Conference on Human Factors in Computing Systems}}. \bibinfo{pages}{1--13}.
\newblock


\bibitem[Davis and Hall(2017)]%
        {davis2017makeup}
\bibfield{author}{\bibinfo{person}{Gretchen Davis} {and} \bibinfo{person}{Mindy Hall}.} \bibinfo{year}{2017}\natexlab{}.
\newblock \bibinfo{booktitle}{\emph{The makeup artist handbook: techniques for film, television, photography, and theatre}}.
\newblock \bibinfo{publisher}{Routledge}.
\newblock


\bibitem[Dellinger and Williams(1997)]%
        {dellinger1997makeup}
\bibfield{author}{\bibinfo{person}{Kirsten Dellinger} {and} \bibinfo{person}{Christine~L Williams}.} \bibinfo{year}{1997}\natexlab{}.
\newblock \showarticletitle{Makeup at work: Negotiating appearance rules in the workplace}.
\newblock \bibinfo{journal}{\emph{Gender \& Society}} \bibinfo{volume}{11}, \bibinfo{number}{2} (\bibinfo{year}{1997}), \bibinfo{pages}{151--177}.
\newblock


\bibitem[Food et~al\mbox{.}(2013)]%
        {us2013draft}
\bibfield{author}{\bibinfo{person}{US Food}, \bibinfo{person}{Drug Administration}, {et~al\mbox{.}}} \bibinfo{year}{2013}\natexlab{}.
\newblock \showarticletitle{Draft guidance for industry: cosmetic good manufacturing practices}.
\newblock  (\bibinfo{year}{2013}).
\newblock


\bibitem[Guo et~al\mbox{.}(2016)]%
        {guo2016vizlens}
\bibfield{author}{\bibinfo{person}{Anhong Guo}, \bibinfo{person}{Xiang'Anthony' Chen}, \bibinfo{person}{Haoran Qi}, \bibinfo{person}{Samuel White}, \bibinfo{person}{Suman Ghosh}, \bibinfo{person}{Chieko Asakawa}, {and} \bibinfo{person}{Jeffrey~P Bigham}.} \bibinfo{year}{2016}\natexlab{}.
\newblock \showarticletitle{Vizlens: A robust and interactive screen reader for interfaces in the real world}. In \bibinfo{booktitle}{\emph{Proceedings of the 29th annual symposium on user interface software and technology}}. \bibinfo{pages}{651--664}.
\newblock


\bibitem[Guo et~al\mbox{.}(2017)]%
        {guo2017facade}
\bibfield{author}{\bibinfo{person}{Anhong Guo}, \bibinfo{person}{Jeeeun Kim}, \bibinfo{person}{Xiang'Anthony' Chen}, \bibinfo{person}{Tom Yeh}, \bibinfo{person}{Scott~E Hudson}, \bibinfo{person}{Jennifer Mankoff}, {and} \bibinfo{person}{Jeffrey~P Bigham}.} \bibinfo{year}{2017}\natexlab{}.
\newblock \showarticletitle{Facade: Auto-generating tactile interfaces to appliances}. In \bibinfo{booktitle}{\emph{Proceedings of the 2017 CHI Conference on Human Factors in Computing Systems}}. \bibinfo{pages}{5826--5838}.
\newblock


\bibitem[He et~al\mbox{.}(2017)]%
        {he2017tactile}
\bibfield{author}{\bibinfo{person}{Liang He}, \bibinfo{person}{Zijian Wan}, \bibinfo{person}{Leah Findlater}, {and} \bibinfo{person}{Jon~E Froehlich}.} \bibinfo{year}{2017}\natexlab{}.
\newblock \showarticletitle{TacTILE: a preliminary toolchain for creating accessible graphics with 3D-printed overlays and auditory annotations}. In \bibinfo{booktitle}{\emph{Proceedings of the 19th International ACM SIGACCESS Conference on Computers and Accessibility}}. \bibinfo{pages}{397--398}.
\newblock


\bibitem[Holly(2018)]%
        {BlindGir58:online}
\bibfield{author}{\bibinfo{person}{Holly}.} \bibinfo{year}{2018}\natexlab{}.
\newblock \bibinfo{title}{Blind Girl Make-up Tips and Tricks - Life of a Blind Girl}.
\newblock \bibinfo{howpublished}{\url{https://lifeofablindgirl.com/2018/07/29/blind-girl-make-up-tips-and-tricks/}}.
\newblock
\newblock
\shownote{(Accessed on 09/07/2021)}.


\bibitem[Inc({[n.\,d.]})]%
        {EstéeLau26:online}
\bibfield{author}{\bibinfo{person}{Estée~Lauder Inc}.} \bibinfo{year}{[n.\,d.]}\natexlab{}.
\newblock \bibinfo{title}{Estée Lauder Voice-enabled Makeup Assistant}.
\newblock
\urldef\tempurl%
\url{https://www.esteelauder.com/voice-enabled-makeup-assistant.tmpl?srsltid=AfmBOopCHyM73CABLtbGIYMNbgDRi-Fpo6UjZBiZ-ADfmJMvCrZw4awk}
\showURL{%
\tempurl}
\newblock
\shownote{[Online; accessed 2025-03-30]}.


\bibitem[Jain and Bhatti(2009)]%
        {jain2009imaging}
\bibfield{author}{\bibinfo{person}{Jhilmil Jain} {and} \bibinfo{person}{Nina Bhatti}.} \bibinfo{year}{2009}\natexlab{}.
\newblock \showarticletitle{Imaging-based cosmetics advisory service}.
\newblock In \bibinfo{booktitle}{\emph{CHI'09 Extended Abstracts on Human Factors in Computing Systems}}. \bibinfo{pages}{3781--3786}.
\newblock


\bibitem[Jairath and Daima({[n.\,d.]})]%
        {jairath1role}
\bibfield{author}{\bibinfo{person}{Jaanvi Jairath} {and} \bibinfo{person}{Rhea Daima}.} \bibinfo{year}{[n.\,d.]}\natexlab{}.
\newblock \showarticletitle{Role of Pop Culture in Popularizing Gender-Bending Fashion and Ideals of Beauty and Makeup}.
\newblock \bibinfo{journal}{\emph{International Journal}} \bibinfo{volume}{1}, \bibinfo{number}{3} (\bibinfo{year}{[n.\,d.]}).
\newblock


\bibitem[Kamikubo et~al\mbox{.}(2024)]%
        {kamikubo2024we}
\bibfield{author}{\bibinfo{person}{Rie Kamikubo}, \bibinfo{person}{Hernisa Kacorri}, {and} \bibinfo{person}{Chieko Asakawa}.} \bibinfo{year}{2024}\natexlab{}.
\newblock \showarticletitle{" We are at the mercy of others' opinion": Supporting Blind People in Recreational Window Shopping with AI-infused Technology}. In \bibinfo{booktitle}{\emph{Proceedings of the 21st International Web for All Conference}}. \bibinfo{pages}{45--58}.
\newblock


\bibitem[Kao et~al\mbox{.}(2016)]%
        {kao2016chromoskin}
\bibfield{author}{\bibinfo{person}{Hsin-Liu Kao}, \bibinfo{person}{Manisha Mohan}, \bibinfo{person}{Chris Schmandt}, \bibinfo{person}{Joseph~A Paradiso}, {and} \bibinfo{person}{Katia Vega}.} \bibinfo{year}{2016}\natexlab{}.
\newblock \showarticletitle{Chromoskin: Towards interactive cosmetics using thermochromic pigments}. In \bibinfo{booktitle}{\emph{Proceedings of the 2016 CHI Conference Extended Abstracts on Human Factors in Computing Systems}}. \bibinfo{pages}{3703--3706}.
\newblock


\bibitem[Kennedy(2016)]%
        {kennedy2016exploring}
\bibfield{author}{\bibinfo{person}{{\"U}mit Kennedy}.} \bibinfo{year}{2016}\natexlab{}.
\newblock \showarticletitle{Exploring YouTube as a transformative tool in the “The Power of MAKEUP!” movement}.
\newblock \bibinfo{journal}{\emph{M/C Journal}} \bibinfo{volume}{19}, \bibinfo{number}{4} (\bibinfo{year}{2016}).
\newblock


\bibitem[Kianpisheh et~al\mbox{.}(2019)]%
        {kianpisheh2019face}
\bibfield{author}{\bibinfo{person}{Mohammad Kianpisheh}, \bibinfo{person}{Franklin~Mingzhe Li}, {and} \bibinfo{person}{Khai~N Truong}.} \bibinfo{year}{2019}\natexlab{}.
\newblock \showarticletitle{Face recognition assistant for people with visual impairments}.
\newblock \bibinfo{journal}{\emph{Proceedings of the ACM on Interactive, Mobile, Wearable and Ubiquitous Technologies}} \bibinfo{volume}{3}, \bibinfo{number}{3} (\bibinfo{year}{2019}), \bibinfo{pages}{1--24}.
\newblock


\bibitem[Korichi and Pelle-de Queral(2008)]%
        {korichi2008women}
\bibfield{author}{\bibinfo{person}{Rodolphe Korichi} {and} \bibinfo{person}{Delphine Pelle-de Queral}.} \bibinfo{year}{2008}\natexlab{}.
\newblock \showarticletitle{Why women use makeup: Implication of psychological}.
\newblock \bibinfo{journal}{\emph{Journal of cosmetic science}}  \bibinfo{volume}{59} (\bibinfo{year}{2008}), \bibinfo{pages}{127--137}.
\newblock


\bibitem[Kr{\'o}lak et~al\mbox{.}(2017)]%
        {krolak2017accessibility}
\bibfield{author}{\bibinfo{person}{Aleksandra Kr{\'o}lak}, \bibinfo{person}{Weiqin Chen}, \bibinfo{person}{Norun~C Sanderson}, {and} \bibinfo{person}{Siri Kessel}.} \bibinfo{year}{2017}\natexlab{}.
\newblock \showarticletitle{The accessibility of MOOCs for blind learners}. In \bibinfo{booktitle}{\emph{Proceedings of the 19th International ACM SIGACCESS Conference on Computers and Accessibility}}. \bibinfo{pages}{401--402}.
\newblock


\bibitem[Lauder({[n.\,d.]})]%
        {EstéeLau38:online}
\bibfield{author}{\bibinfo{person}{Estée Lauder}.} \bibinfo{year}{[n.\,d.]}\natexlab{}.
\newblock \bibinfo{title}{Estée Lauder Voice-enabled Makeup Assistant}.
\newblock
\urldef\tempurl%
\url{https://www.esteelauder.com/voice-enabled-makeup-assistant.tmpl?srsltid=AfmBOornwC4Mb9R6uddDMtDyYGsZt7PQ3mKkTpviGmePGMvd0O5ANgyP}
\showURL{%
\tempurl}
\newblock
\shownote{[Online; accessed 2025-04-16]}.


\bibitem[Li et~al\mbox{.}(2021a)]%
        {li2021choose}
\bibfield{author}{\bibinfo{person}{Franklin~Mingzhe Li}, \bibinfo{person}{Di~Laura Chen}, \bibinfo{person}{Mingming Fan}, {and} \bibinfo{person}{Khai~N Truong}.} \bibinfo{year}{2021}\natexlab{a}.
\newblock \showarticletitle{“I Choose Assistive Devices That Save My Face” A Study on Perceptions of Accessibility and Assistive Technology Use Conducted in China}. In \bibinfo{booktitle}{\emph{Proceedings of the 2021 CHI Conference on Human Factors in Computing Systems}}. \bibinfo{pages}{1--14}.
\newblock


\bibitem[Li et~al\mbox{.}(2021b)]%
        {li2021non}
\bibfield{author}{\bibinfo{person}{Franklin~Mingzhe Li}, \bibinfo{person}{Jamie Dorst}, \bibinfo{person}{Peter Cederberg}, {and} \bibinfo{person}{Patrick Carrington}.} \bibinfo{year}{2021}\natexlab{b}.
\newblock \showarticletitle{Non-Visual Cooking: Exploring Practices and Challenges of Meal Preparation by People with Visual Impairments}. In \bibinfo{booktitle}{\emph{The 23rd International ACM SIGACCESS Conference on Computers and Accessibility}}. \bibinfo{pages}{1--11}.
\newblock


\bibitem[Li et~al\mbox{.}(2024a)]%
        {li2024contextual}
\bibfield{author}{\bibinfo{person}{Franklin~Mingzhe Li}, \bibinfo{person}{Michael~Xieyang Liu}, \bibinfo{person}{Shaun~K Kane}, {and} \bibinfo{person}{Patrick Carrington}.} \bibinfo{year}{2024}\natexlab{a}.
\newblock \showarticletitle{A Contextual Inquiry of People with Vision Impairments in Cooking}. In \bibinfo{booktitle}{\emph{Proceedings of the 2024 CHI Conference on Human Factors in Computing Systems}}. \bibinfo{pages}{1--14}.
\newblock


\bibitem[Li et~al\mbox{.}(2025)]%
        {li2025oscar}
\bibfield{author}{\bibinfo{person}{Franklin~Mingzhe Li}, \bibinfo{person}{Kaitlyn Ng}, \bibinfo{person}{Bin Zhu}, {and} \bibinfo{person}{Patrick Carrington}.} \bibinfo{year}{2025}\natexlab{}.
\newblock \showarticletitle{OSCAR: Object Status and Contextual Awareness for Recipes to Support Non-Visual Cooking}. In \bibinfo{booktitle}{\emph{Proceedings of the Extended Abstracts of the CHI Conference on Human Factors in Computing Systems}}. \bibinfo{pages}{1--6}.
\newblock


\bibitem[Li et~al\mbox{.}(2022)]%
        {li2022feels}
\bibfield{author}{\bibinfo{person}{Franklin~Mingzhe Li}, \bibinfo{person}{Franchesca Spektor}, \bibinfo{person}{Meng Xia}, \bibinfo{person}{Mina Huh}, \bibinfo{person}{Peter Cederberg}, \bibinfo{person}{Yuqi Gong}, \bibinfo{person}{Kristen Shinohara}, {and} \bibinfo{person}{Patrick Carrington}.} \bibinfo{year}{2022}\natexlab{}.
\newblock \showarticletitle{“It feels like taking a gamble”: Exploring perceptions, practices, and challenges of using makeup and cosmetics for people with visual impairments}. In \bibinfo{booktitle}{\emph{Proceedings of the 2022 CHI Conference on Human Factors in Computing Systems}}. \bibinfo{pages}{1--15}.
\newblock


\bibitem[Li et~al\mbox{.}(2024b)]%
        {li2024recipe}
\bibfield{author}{\bibinfo{person}{Franklin~Mingzhe Li}, \bibinfo{person}{Ashley Wang}, \bibinfo{person}{Patrick Carrington}, {and} \bibinfo{person}{Shaun~K Kane}.} \bibinfo{year}{2024}\natexlab{b}.
\newblock \showarticletitle{A Recipe for Success? Exploring Strategies for Improving Non-Visual Access to Cooking Instructions}. In \bibinfo{booktitle}{\emph{Proceedings of the 26th International ACM SIGACCESS Conference on Computers and Accessibility}}. \bibinfo{pages}{1--15}.
\newblock


\bibitem[Li et~al\mbox{.}(2023)]%
        {li2023understanding}
\bibfield{author}{\bibinfo{person}{Franklin~Mingzhe Li}, \bibinfo{person}{Lotus Zhang}, \bibinfo{person}{Maryam Bandukda}, \bibinfo{person}{Abigale Stangl}, \bibinfo{person}{Kristen Shinohara}, \bibinfo{person}{Leah Findlater}, {and} \bibinfo{person}{Patrick Carrington}.} \bibinfo{year}{2023}\natexlab{}.
\newblock \showarticletitle{Understanding visual arts experiences of blind people}. In \bibinfo{booktitle}{\emph{Proceedings of the 2023 CHI Conference on Human Factors in Computing Systems}}. \bibinfo{pages}{1--21}.
\newblock


\bibitem[Li et~al\mbox{.}(2017)]%
        {li2017braillesketch}
\bibfield{author}{\bibinfo{person}{Mingzhe Li}, \bibinfo{person}{Mingming Fan}, {and} \bibinfo{person}{Khai~N Truong}.} \bibinfo{year}{2017}\natexlab{}.
\newblock \showarticletitle{BrailleSketch: A gesture-based text input method for people with visual impairments}. In \bibinfo{booktitle}{\emph{Proceedings of the 19th International ACM SIGACCESS Conference on Computers and Accessibility}}. \bibinfo{pages}{12--21}.
\newblock


\bibitem[Liu et~al\mbox{.}(2021)]%
        {liu2021tactile}
\bibfield{author}{\bibinfo{person}{Guanhong Liu}, \bibinfo{person}{Tianyu Yu}, \bibinfo{person}{Chun Yu}, \bibinfo{person}{Haiqing Xu}, \bibinfo{person}{Shuchang Xu}, \bibinfo{person}{Ciyuan Yang}, \bibinfo{person}{Feng Wang}, \bibinfo{person}{Haipeng Mi}, {and} \bibinfo{person}{Yuanchun Shi}.} \bibinfo{year}{2021}\natexlab{}.
\newblock \showarticletitle{Tactile compass: Enabling visually impaired people to follow a path with continuous directional feedback}. In \bibinfo{booktitle}{\emph{Proceedings of the 2021 CHI Conference on Human Factors in Computing Systems}}.
\newblock


\bibitem[Long(2019)]%
        {Blindnes56:online}
\bibfield{author}{\bibinfo{person}{April Long}.} \bibinfo{year}{2019}\natexlab{}.
\newblock \bibinfo{title}{Blindness \& Beauty: How Visually Impaired Women Are Changing the Industry | Allure}.
\newblock \bibinfo{howpublished}{\url{https://www.allure.com/story/blind-women-beauty-industry-tactile-packaging-for-visually-impaired}}.
\newblock
\newblock
\shownote{(Accessed on 09/07/2021)}.


\bibitem[Mascetti et~al\mbox{.}(2016)]%
        {mascetti2016towards}
\bibfield{author}{\bibinfo{person}{Sergio Mascetti}, \bibinfo{person}{Chiara Rossetti}, \bibinfo{person}{Andrea Gerino}, \bibinfo{person}{Cristian Bernareggi}, \bibinfo{person}{Lorenzo Picinali}, {and} \bibinfo{person}{Alessandro Rizzi}.} \bibinfo{year}{2016}\natexlab{}.
\newblock \showarticletitle{Towards a natural user interface to support people with visual impairments in detecting colors}. In \bibinfo{booktitle}{\emph{International Conference on Computers Helping People with Special Needs}}. Springer, \bibinfo{pages}{171--178}.
\newblock


\bibitem[Neiva(2017)]%
        {neiva2017coloradd}
\bibfield{author}{\bibinfo{person}{Miguel Neiva}.} \bibinfo{year}{2017}\natexlab{}.
\newblock \showarticletitle{ColorADD: color identification system for color-blind people}.
\newblock In \bibinfo{booktitle}{\emph{Injuries and Health Problems in Football}}. \bibinfo{publisher}{Springer}, \bibinfo{pages}{303--314}.
\newblock


\bibitem[Nguyen and Liu(2017)]%
        {nguyen2017smart}
\bibfield{author}{\bibinfo{person}{Tam~V Nguyen} {and} \bibinfo{person}{Luoqi Liu}.} \bibinfo{year}{2017}\natexlab{}.
\newblock \showarticletitle{Smart mirror: Intelligent makeup recommendation and synthesis}. In \bibinfo{booktitle}{\emph{Proceedings of the 25th ACM international conference on Multimedia}}. \bibinfo{pages}{1253--1254}.
\newblock


\bibitem[Ou et~al\mbox{.}(2016)]%
        {ou2016beauty}
\bibfield{author}{\bibinfo{person}{Xinyu Ou}, \bibinfo{person}{Si Liu}, \bibinfo{person}{Xiaochun Cao}, {and} \bibinfo{person}{Hefei Ling}.} \bibinfo{year}{2016}\natexlab{}.
\newblock \showarticletitle{Beauty emakeup: A deep makeup transfer system}. In \bibinfo{booktitle}{\emph{Proceedings of the 24th ACM international conference on Multimedia}}. \bibinfo{pages}{701--702}.
\newblock


\bibitem[Pradhan et~al\mbox{.}(2021)]%
        {pradhan2021inclusive}
\bibfield{author}{\bibinfo{person}{Akriti Pradhan}, \bibinfo{person}{Gabriela Daniels}, {et~al\mbox{.}}} \bibinfo{year}{2021}\natexlab{}.
\newblock \showarticletitle{Inclusive beauty: how buying and using cosmetics can be made more accessible for the visually impaired (VI) and blind consumer}.
\newblock \bibinfo{journal}{\emph{Cosmetics and Toiletries}} \bibinfo{volume}{136}, \bibinfo{number}{4} (\bibinfo{year}{2021}), \bibinfo{pages}{DM4--DM15}.
\newblock


\bibitem[Quito(2018)]%
        {Inclusiv52:online}
\bibfield{author}{\bibinfo{person}{Anne Quito}.} \bibinfo{year}{2018}\natexlab{}.
\newblock \bibinfo{title}{Inclusive design: Procter \& Gamble updates Herbal Essences shampoo bottles for blind customers — Quartz}.
\newblock \bibinfo{howpublished}{\url{https://qz.com/quartzy/1418531/inclusive-design-procter-gamble-updates-herbal-essences-shampoo-bottles-for-blind-customers/}}.
\newblock
\newblock
\shownote{(Accessed on 08/10/2021)}.


\bibitem[Ran et~al\mbox{.}(2025)]%
        {ran2025users}
\bibfield{author}{\bibinfo{person}{Zihe Ran}, \bibinfo{person}{Xiyu Li}, \bibinfo{person}{Qing Xiao}, \bibinfo{person}{Xianzhe Fan}, \bibinfo{person}{Franklin~Mingzhe Li}, \bibinfo{person}{Yanyun Wang}, {and} \bibinfo{person}{Zhicong Lu}.} \bibinfo{year}{2025}\natexlab{}.
\newblock \showarticletitle{How Users Who are Blind or Low Vision Play Mobile Games: Perceptions, Challenges, and Strategies}. In \bibinfo{booktitle}{\emph{Proceedings of the 2025 CHI Conference on Human Factors in Computing Systems}}. \bibinfo{pages}{1--18}.
\newblock


\bibitem[Sato et~al\mbox{.}(2017)]%
        {sato2017navcog3}
\bibfield{author}{\bibinfo{person}{Daisuke Sato}, \bibinfo{person}{Uran Oh}, \bibinfo{person}{Kakuya Naito}, \bibinfo{person}{Hironobu Takagi}, \bibinfo{person}{Kris Kitani}, {and} \bibinfo{person}{Chieko Asakawa}.} \bibinfo{year}{2017}\natexlab{}.
\newblock \showarticletitle{Navcog3: An evaluation of a smartphone-based blind indoor navigation assistant with semantic features in a large-scale environment}. In \bibinfo{booktitle}{\emph{Proceedings of the 19th International ACM SIGACCESS Conference on Computers and Accessibility}}. \bibinfo{pages}{270--279}.
\newblock


\bibitem[Seo and Jung(2018)]%
        {seo2018understanding}
\bibfield{author}{\bibinfo{person}{Woosuk Seo} {and} \bibinfo{person}{Hyunggu Jung}.} \bibinfo{year}{2018}\natexlab{}.
\newblock \showarticletitle{Understanding blind or visually impaired people on youtube through qualitative analysis of videos}. In \bibinfo{booktitle}{\emph{Proceedings of the 2018 ACM International Conference on Interactive Experiences for TV and Online Video}}. \bibinfo{pages}{191--196}.
\newblock


\bibitem[Shinohara and Wobbrock(2011)]%
        {shinohara2011shadow}
\bibfield{author}{\bibinfo{person}{Kristen Shinohara} {and} \bibinfo{person}{Jacob~O Wobbrock}.} \bibinfo{year}{2011}\natexlab{}.
\newblock \showarticletitle{In the shadow of misperception: assistive technology use and social interactions}. In \bibinfo{booktitle}{\emph{Proceedings of the SIGCHI conference on human factors in computing systems}}. \bibinfo{pages}{705--714}.
\newblock


\bibitem[Sicardi(2019)]%
        {Beautyis6:online}
\bibfield{author}{\bibinfo{person}{Arabelle Sicardi}.} \bibinfo{year}{2019}\natexlab{}.
\newblock \bibinfo{title}{Beauty is designing packaging for the visually impaired | Vogue Business}.
\newblock \bibinfo{howpublished}{\url{https://www.voguebusiness.com/beauty/braille-beauty-packaging-loccitane}}.
\newblock
\newblock
\shownote{(Accessed on 06/14/2021)}.


\bibitem[Treepong et~al\mbox{.}(2018)]%
        {treepong2018makeup}
\bibfield{author}{\bibinfo{person}{Bantita Treepong}, \bibinfo{person}{Hironori Mitake}, {and} \bibinfo{person}{Shoichi Hasegawa}.} \bibinfo{year}{2018}\natexlab{}.
\newblock \showarticletitle{Makeup Creativity Enhancement with an Augmented Reality Face Makeup System}.
\newblock \bibinfo{journal}{\emph{Computers in Entertainment (CIE)}} \bibinfo{volume}{16}, \bibinfo{number}{4} (\bibinfo{year}{2018}), \bibinfo{pages}{1--17}.
\newblock


\bibitem[Truong et~al\mbox{.}(2021)]%
        {truong2021automatic}
\bibfield{author}{\bibinfo{person}{Anh Truong}, \bibinfo{person}{Peggy Chi}, \bibinfo{person}{David Salesin}, \bibinfo{person}{Irfan Essa}, {and} \bibinfo{person}{Maneesh Agrawala}.} \bibinfo{year}{2021}\natexlab{}.
\newblock \showarticletitle{Automatic Generation of Two-Level Hierarchical Tutorials from Instructional Makeup Videos}. In \bibinfo{booktitle}{\emph{Proceedings of the 2021 CHI Conference on Human Factors in Computing Systems}}. \bibinfo{pages}{1--16}.
\newblock


\bibitem[Wang et~al\mbox{.}(2021)]%
        {wang2021revamp}
\bibfield{author}{\bibinfo{person}{Ruolin Wang}, \bibinfo{person}{Zixuan Chen}, \bibinfo{person}{Mingrui~Ray Zhang}, \bibinfo{person}{Zhaoheng Li}, \bibinfo{person}{Zhixiu Liu}, \bibinfo{person}{Zihan Dang}, \bibinfo{person}{Chun Yu}, {and} \bibinfo{person}{Xiang'Anthony' Chen}.} \bibinfo{year}{2021}\natexlab{}.
\newblock \showarticletitle{Revamp: Enhancing Accessible Information Seeking Experience of Online Shopping for Blind or Low Vision Users}. In \bibinfo{booktitle}{\emph{Proceedings of the 2021 CHI Conference on Human Factors in Computing Systems}}. \bibinfo{pages}{1--14}.
\newblock


\bibitem[WHO(2021)]%
        {Blindnes59:online}
\bibfield{author}{\bibinfo{person}{WHO}.} \bibinfo{year}{2021}\natexlab{}.
\newblock \bibinfo{title}{Blindness and vision impairment}.
\newblock \bibinfo{howpublished}{\url{https://www.who.int/news-room/fact-sheets/detail/blindness-and-visual-impairment}}.
\newblock
\newblock
\shownote{(Accessed on 06/14/2021)}.


\bibitem[Yatani et~al\mbox{.}(2012)]%
        {yatani2012spacesense}
\bibfield{author}{\bibinfo{person}{Koji Yatani}, \bibinfo{person}{Nikola Banovic}, {and} \bibinfo{person}{Khai Truong}.} \bibinfo{year}{2012}\natexlab{}.
\newblock \showarticletitle{SpaceSense: representing geographical information to visually impaired people using spatial tactile feedback}. In \bibinfo{booktitle}{\emph{Proceedings of the SIGCHI Conference on Human Factors in Computing Systems}}. \bibinfo{pages}{415--424}.
\newblock


\end{thebibliography}


\end{document}